%% file: main.tex
\PassOptionsToPackage{unicode}{hyperref}
\PassOptionsToPackage{hyphens}{url}
\PassOptionsToPackage{dvipsnames,svgnames,x11names}{xcolor}
\documentclass[letterpaper,10pt,twocolumn]{article}
\usepackage{xcolor}

\usepackage[
  letterpaper,
  textwidth=7in,
  textheight=9in,
  centering,
  columnsep=0.33in,
  headheight=14pt,
]{geometry}

\usepackage{amsmath,amssymb,amsthm,mathtools}
\usepackage{newunicodechar}
\usepackage{fvextra}
\fvset{breaklines=true,breakanywhere=true,fontsize=\footnotesize,xleftmargin=0pt}
\newunicodechar{↾}{\ensuremath{\upharpoonright}}

\newcommand{\restrict}{\mathbin{\upharpoonright}}
\newcommand{\Apply}{\mathit{apply}}
\newcommand{\Src}{\mathit{src}}
\newcommand{\Scope}{\mathit{scope}}
\newcommand{\Frontier}{\mathit{frontier}}
\newcommand{\CleanPrefix}{\mathit{prefix}}
\newcommand{\Verify}{\mathit{verify}}
\newcommand{\AnchorDomain}{\mathit{anchorDom}}
\newcommand{\TouchedBetween}{\mathit{touched}}
\newcommand{\TableScopeBetween}{\mathit{tblScope}}
\newcommand{\ReplayKeys}{\mathit{keys}}
\newcommand{\Chunks}{\mathit{chunks}}
\newcommand{\SourceHistory}{\mathit{srcHistory}}
\newcommand{\CdcEvents}{\mathit{cdcEvents}}
\newcommand{\RawObs}{\mathit{raw}}
\newcommand{\SinkState}{\mathit{sinkState}}
\newcommand{\WellformedRun}{\mathsf{WF}}
\newcommand{\WF}{\mathsf{WF}}
\newcommand{\FaithfulSourceObservation}{\mathsf{FSO}}
\newcommand{\FSO}{\mathsf{FSO}}

\newcommand{\Mat}{\mathsf{Mat}}

\newcommand{\Coh}{\mathsf{Coh}}
\newcommand{\VirtualCut}{\mathsf{VC}}
\newcommand{\VC}{\mathsf{VC}}

\newcommand{\VCS}{\mathsf{VCS}}
\newcommand{\WholeTableClaimScope}{\mathsf{WTS}}
\newcommand{\WTS}{\mathsf{WTS}}
\newcommand{\ContSeg}{\mathsf{ContSeg}}
\newcommand{\Accept}{\mathsf{Accept}}
\newcommand{\Reject}{\mathsf{Reject}}
\newcommand{\Unsupported}{\mathsf{Unsupported}}
\newcommand{\None}{\mathrm{None}}

\usepackage{iftex}
\ifPDFTeX
  \usepackage[T1]{fontenc}
  \usepackage[utf8]{inputenc}
  \usepackage{textcomp}
  \IfFileExists{newtxtext.sty}{%
    \usepackage{newtxtext}%
    \usepackage[varvw]{newtxmath}%
  }{\usepackage{lmodern}}
\else
  \usepackage{unicode-math}
  \defaultfontfeatures{Scale=MatchLowercase}
  \defaultfontfeatures[\rmfamily]{Ligatures=TeX,Scale=1}
  \usepackage{lmodern}
\fi
\IfFileExists{upquote.sty}{\usepackage{upquote}}{}
\IfFileExists{microtype.sty}{%
  \usepackage{microtype}%
  \UseMicrotypeSet[protrusion]{basicmath}%
}{}

\usepackage{booktabs,array,tabularx,ragged2e}
\newcolumntype{Y}{>{\RaggedRight\arraybackslash}X}
\providecommand{\tightlist}{\setlength{\itemsep}{0pt}\setlength{\parskip}{0pt}}

\usepackage{enumitem}
\setlist{topsep=2pt,partopsep=0pt,parsep=2pt,itemsep=2pt,leftmargin=*}

\usepackage[numbers,sort&compress]{natbib}

\usepackage{placeins}

\usepackage{sectsty}
\allsectionsfont{\raggedright\hyphenpenalty=10000\exhyphenpenalty=10000}

\usepackage{bookmark}
\IfFileExists{xurl.sty}{\usepackage{xurl}}{}
\hypersetup{
  pdftitle={A Theoretical Study of DBLog: Certified Virtual Cuts for a Snapshot-Equivalent Replay of Live Databases},
  pdfauthor={Andreas Andreakis},
  hidelinks}

\usepackage[normalem]{ulem}
\usepackage{tikz}
\usetikzlibrary{arrows.meta,decorations.pathreplacing,decorations.markings,%
                positioning,calc,shapes.geometric,shapes.misc,fit,%
                backgrounds,patterns,matrix,quotes}

\definecolor{gDark}{HTML}{3A3A3A}
\definecolor{gMid}{HTML}{6A6A6A}
\definecolor{gLight}{HTML}{A6A6A6}
\definecolor{gPale}{HTML}{E5E5E5}
\definecolor{gBg}{HTML}{F4F4F4}

\tikzset{
  setA/.style={
    font=\sffamily\footnotesize,
    every node/.style={inner sep=2.5pt, outer sep=0pt},
    >={Latex[length=4pt, width=3pt]},
    line cap=round, line join=round,
    line width=0.5pt,
  },
  block/.style={
    draw=gDark, line width=0.5pt, rounded corners=1pt,
    fill=white, align=center, inner sep=4pt,
  },
  blockFill/.style={block, fill=gPale},
  blockEmph/.style={block, line width=0.8pt},
  blockGhost/.style={
    draw=gLight, line width=0.4pt, rounded corners=1pt,
    fill=white, dashed, align=center, inner sep=4pt, text=gMid,
  },
  layerChip/.style={
    draw=gDark, line width=0.5pt, rounded corners=2pt,
    fill=gPale, minimum width=11mm, minimum height=8mm,
    font=\sffamily\bfseries\small, align=center, inner sep=2pt,
  },
  layerChipGhost/.style={
    layerChip, draw=gLight, fill=white, text=gMid,
    font=\sffamily\small,
  },
  axisLine/.style={-{Latex[length=3.2pt, width=2.4pt]}, gDark, line width=0.45pt},
  axisTick/.style={gDark, line width=0.45pt},
  flow/.style={-{Latex[length=3.5pt, width=2.6pt]}, gDark, line width=0.45pt},
  depFlow/.style={-{Latex[length=3.5pt, width=2.6pt]}, gMid, line width=0.45pt},
  weakFlow/.style={-{Latex[length=3.5pt, width=2.6pt]}, gLight, line width=0.4pt, dashed},
  guide/.style={gMid, line width=0.4pt, dotted},
  cutLine/.style={gDark, line width=1.4pt},
  brace/.style={decorate, decoration={brace, amplitude=3pt, raise=1pt}, gDark, line width=0.4pt},
  braceMirror/.style={decorate, decoration={brace, mirror, amplitude=3pt, raise=1pt}, gDark, line width=0.4pt},
  pill/.style={
    draw=gDark, line width=0.4pt, rounded corners=2pt,
    fill=white, inner xsep=3pt, inner ysep=1.5pt, align=center,
    font=\sffamily\scriptsize,
  },
  pillFill/.style={pill, fill=gPale},
  lane/.style={gMid, line width=0.35pt},
  sublabel/.style={font=\sffamily\scriptsize, text=gDark},
  sublabelDim/.style={font=\sffamily\scriptsize\itshape, text=gMid},
}

\title{A Theoretical Study of DBLog: Certified Virtual Cuts for a Snapshot-Equivalent Replay of Live Databases}
\author{Andreas Andreakis}

\makeatletter
\renewcommand{\maketitle}{%
  \twocolumn[{%
    \begin{center}
      {\LARGE\bfseries \@title\par}
      \vspace{1.1em}
      {\large \@author\par}
    \end{center}
    \vspace{1.5em}
  }]%
  \thispagestyle{plain}%
}
\makeatother
\date{}

\begin{document}
\maketitle

\section*{Abstract}\label{abstract}

DBLog is a change-data-capture (CDC) mechanism for copying a table or
selected keys from a source database while continuing to stream new changes
from its commit log. It reads a table in primary-key ordered chunks and 
brackets each read with low and high watermarks in the source log. A chunk
row is discarded if a log event for the same key appears anywhere in that
window. The remaining rows are emitted as refresh events after the high
watermark. DBLog allows backfills to run at any time during normal operation,
rather than only as an initial bootstrap, while writes and live capture
continue. It was introduced in the 2019 Netflix Tech Blog
post~\citep{dblog_netflix_blog} and further discussed in the 2020 DBLog
paper~\citep{andreakis2020dblog}. The mechanism has since been adopted by
open-source projects, including Debezium~\citep{debezium_incremental_snapshot}
and Apache Flink CDC~\citep{flink_cdc_incremental_snapshot}.

The original blog post and paper explained this mechanism but did not provide
a formal correctness proof or a precise description of its replay result. In
this paper, we formalize how chunk reads and log events are
combined and prove that their replay reconstructs the source state up to a
specific log position for the keys being copied. We define this as a
\textbf{virtual cut}, representing a snapshot-equivalent replay without
requiring a physical snapshot. DBLog emits the events needed to reconstruct 
the source state downstream, without storing that state itself.

To verify executions independently, we provide a certificate that validates
this equivalence from recorded log evidence and chunk observations. Once chunk 
processing completes and covers all keys, this guarantee extends to the whole table. 
Furthermore, appending subsequent change-stream events advances the cut forward, 
formalizing the seamless transition from historical backfill to live streaming. 
All definitions and proofs are mechanized in Isabelle/HOL.

\section{Introduction}\label{introduction}

DBLog, introduced in the Netflix Tech Blog in 2019~\citep{dblog_netflix_blog} 
and described in the 2020 DBLog paper~\citep{andreakis2020dblog}, is a 
change-data-capture (CDC) mechanism for backfilling a downstream system from a
source database while continuing to stream new changes. DBLog reads the 
table in primary-key chunks and brackets each chunk read between low and high 
watermarks in the source log. The watermarks bracket the read without
knowing its exact source-log position. DBLog then discards any chunk row
whose key collides with a CDC event anywhere within that window to ensure
that refresh events don't override more recent CDC events. CDC events
retain their source order, and the remaining chunk rows are emitted as
refresh events after the high watermark. Variants of this mechanism are now used by
Debezium~\citep{debezium_incremental_snapshot} and Apache Flink 
CDC~\citep{flink_cdc_incremental_snapshot}, where the backfill phase 
is called an \emph{incremental snapshot}.

The mechanism was designed to allow backfills at any time, not only
initially, without interrupting the flow of change events. A backfill 
can run during normal operation, cover an entire table or selected keys, 
and pause or resume between chunks. This lets operators control the 
scope and pace of the work without stalling live capture.

Because chunks are read at distinct source-log positions, intermediate writes 
at the source can render individual row observations stale. The collected chunk reads
 do not form a single physical snapshot. We establish that DBLog's 
combination of chunk refreshes and change events nonetheless reconstructs 
the exact per-key state of the source at a designated commit-log frontier, 
and we prove how that equivalence advances forward as the source commits 
subsequent writes.

While the original 2020 paper explained how the mechanism works operationally, 
it did not formalize the reconstructed database state or provide a proof. 
We supply that foundation by formalizing how chunk observations and captured
changes are combined and proving that their replay reconstructs the source
state. We characterize this result through the concept of a
\textbf{virtual cut}. A
\textbf{virtual cut} formalizes snapshot equivalence: replaying the sequence 
of chunk observations and streaming writes gathered during a backfill produces 
the exact state of the source database at a specific commit-log coordinate, for 
every key in the targeted scope. The cut is \emph{virtual} because rows were read 
at different times without taking a physical snapshot. A \textbf{certified virtual cut} 
packages this claim into an inspectable certificate, allowing an independent checker 
to verify that the replay matches the source from recorded log watermarks and chunk 
reads.

The proof is organized in five layers. Layers~0 and~1 define source 
histories, replay events, and abstract DBLog runs characterized 
by a scope (the targeted set of primary keys) and a frontier 
(the designated commit-log coordinate), alongside the conditions for a 
wellformed run. Layer~2 proves that every wellformed run produces a 
virtual cut at that frontier on that scope. Layer~3 lifts this 
guarantee to inspectable certificates, showing that an independent 
checker can verify the cut directly from recorded log evidence. 
Layer~4 derives full whole-table equality, establishing the 
coverage conditions needed to prove that all existing rows and 
concurrent writes across the table are accounted for. The watermark 
mechanics enter as assumptions, including the single-coordinate 
attribution of each chunk read described in Section~\ref{watermarks-and-chunk-read-coordinates}. 
They are not derived by the abstract model. Figure~\ref{fig:theorem-ladder} 
summarizes these dependencies. Every theorem is machine-checked in the 
companion Isabelle/HOL development, archived at
\url{https://doi.org/10.5281/zenodo.21732790}.

\begin{figure*}[t]
  \input{figures/fig01_theorem_ladder/fig01.tex}
\end{figure*}

The initial certified cut is not the end of DBLog operation. Once the
chunk loop completes, DBLog continues consuming ordinary CDC events. We
prove that appending exactly the source events for keys in scope between
frontiers $f$ and $f'$ advances a cut at $f$ to a cut at $f'$. We also
prove that a cut on one scope remains valid when restricted to a smaller
scope. Normalization of overlapping or retried chunks, and sink state 
correctness remain future work.

The theorems assume that the source log is retained through the claimed frontier, 
each chunk read captures its rows at a single commit-log position, and the CDC 
stream delivers all intermediate change events in committed order. A certificate 
checker verifies that this recorded evidence is internally consistent, but cannot 
determine whether an external connector observed the physical database accurately. 
These guarantees are strictly source-side and scoped to the backfilled keys,
with whole-table equality following under the Layer~4 conditions. 
They do not assert downstream sink convergence or exactly-once delivery.

Section~2 develops a running example and Section~3 reviews the mechanism
used by the model. Section~4 summarizes the claims and limitations.
Sections~5 through~10 define the formal objects and prove the core
results. Section~11 proves continuation and restriction. Section~12
separates source-side replay correctness from destination-state
convergence. Sections~13 through~15 cover related work, future
extensions, and verification methodology.

We use five status labels throughout
(Table~\ref{tab:status-vocab}).

\begin{table*}[t]
\caption{Status vocabulary used throughout the paper.}
\label{tab:status-vocab}
\centering\small
\begin{tabularx}{\textwidth}{@{}lY@{}}
\toprule
\textbf{Status} & \textbf{Meaning in this paper}\\
\midrule
Machine-checked &
  The claim is proved in the companion Isabelle/HOL development, under
  the assumptions stated where the claim appears. The paper gives the proof
  idea or proof explanation in the body.\\
Deployment obligation &
  The operator or implementation must establish this condition in a
  deployment.\\
External observation assumption &
  The source-side observation is assumed accurate (satisfying $\FSO$).
  The model cannot prove it from the certificate alone.\\
Future work &
  A named extension or theorem path that this paper does not claim.\\
Non-claim &
  A property that the paper does not prove.\\
\bottomrule
\end{tabularx}
\end{table*}

Each theorem states its assumptions directly. These labels distinguish the
machine-checked result from conditions supplied by a deployment or its
source connector.

\section{Bootstrapping and Advancing a Virtual Cut with a Running
Example}\label{bootstrapping-and-advancing-a-virtual-cut}

We begin with a concrete backfill before introducing the formal model.
The example shows what DBLog reads while the source accepts concurrent
writes, how those observations form a replay, and why that replay can
match the source without being a physical snapshot. It also shows how
the matching point advances when the source commits another event.

We use this example throughout the paper. This section presents the
mechanism operationally, and Section~5 introduces the corresponding
definitions.

\subsection{Setup}\label{setup}

The source database contains an account table

\begin{center}\ttfamily\small
accounts(account\_id, balance)
\end{center}

initially populated with two rows:

\begin{center}\footnotesize
\begin{tabular}{@{}l@{\hskip 1em}l@{\hskip 1em}l@{}}
\ttfamily account\_id = 100 & \ttfamily balance = \phantom{0}5000 & \itshape (Alice's account, \$50)\\
\ttfamily account\_id = 200 & \ttfamily balance = 10000           & \itshape (Bob's account, \$100)\\
\end{tabular}
\end{center}

A third user, Carol, has not yet opened an account. Therefore,
account\_id 300 does not exist in the table.

During the backfill, application writes continue while DBLog reads
chunks and watches the source log. We construct a replay whose
per-account outcome matches the source at a chosen frontier, although
the chunk reads occur at different source-log coordinates.

\subsection{What the Source Log Records During the Backfill
Window}\label{what-the-source-log-records-during-the-backfill-window}

While DBLog is working, four writes commit in the source log in
the following order:

\begin{center}\footnotesize
\begin{tabular}{@{}l@{\hskip 1em}l@{\hskip 1em}l@{}}
\ttfamily c1 & \ttfamily Update 100 $\to$ \phantom{0}3000 & \itshape (Alice's balance changes to \$30)\\
\ttfamily c2 & \ttfamily Insert 300 $\to$ \phantom{0}2500  & \itshape (Carol opens an account at \$25)\\
\ttfamily c3 & \ttfamily Update 200 $\to$ 12000           & \itshape (Bob's balance changes to \$120)\\
\ttfamily c4 & \ttfamily Update 300 $\to$ \phantom{0}3500 & \itshape (Carol's balance changes to \$35)\\
\end{tabular}
\end{center}

Each $c_i$ denotes a source-log coordinate representing a position in
the source database's commit order, where $c_1 < c_2 < c_3 < c_4$.

After $c_4$ commits, the source state across these three accounts is

\begin{center}\footnotesize
\begin{tabular}{@{}l@{\hskip 1em}l@{\hskip 1em}l@{}}
\ttfamily account 100 & \ttfamily balance = \phantom{0}3000 & \itshape (after the c1 update)\\
\ttfamily account 200 & \ttfamily balance = 12000           & \itshape (after the c3 update)\\
\ttfamily account 300 & \ttfamily balance = \phantom{0}3500 & \itshape (after the c4 update)\\
\end{tabular}
\end{center}

The assembled replay must reconstruct this per-account state.

\subsection{What the Chunks Read}\label{what-the-chunks-read}

DBLog reads the table while source writes continue. It partitions the table
into two \textbf{chunks} by primary key range:

\begin{itemize}
\tightlist
\item
  \textbf{Chunk A} covers $\mathit{account\_id} \in \{100, 200\}$.
\item
  \textbf{Chunk B} covers $\mathit{account\_id} \in \{300\}$.
\end{itemize}

The two chunk ranges are disjoint, and together they cover every
account the backfill claims. Key partitioning is configured by the
deployment rather than inferred by DBLog. Chunks can also cover keys
that do not yet exist in the table. For instance, chunk B covers
account 300 before Carol opens it.

Each chunk is read at its own moment in source-log time. The
deployment must ensure that a multi-row chunk read reflects a single
source-log coordinate, an obligation made precise in
Section~\ref{watermarks-and-chunk-read-coordinates}.

\textbf{Chunk A} runs at coordinate $c_1$, observing
the state immediately after $c_1$'s update committed:

\begin{center}\footnotesize
\begin{tabular}{@{}l@{\hskip 1em}l@{\hskip 1em}l@{}}
\multicolumn{3}{@{}l}{\ttfamily chunk A read at c1:}\\
\ttfamily \hspace{0.5em}account 100 & \ttfamily balance = \phantom{0}3000 & \itshape (already updated by c1)\\
\ttfamily \hspace{0.5em}account 200 & \ttfamily balance = 10000           & \itshape (no write to 200 yet)\\
\end{tabular}
\end{center}

\textbf{Chunk B} runs later, at coordinate $c_3$, observing
Carol's row after Carol's $c_2$ insert but before
Carol's $c_4$ update:

\begin{center}\footnotesize
\begin{tabular}{@{}l@{\hskip 1em}l@{\hskip 1em}l@{}}
\multicolumn{3}{@{}l}{\ttfamily chunk B read at c3:}\\
\ttfamily \hspace{0.5em}account 300 & \ttfamily balance = 2500 & \itshape (Carol's initial balance from c2)\\
\end{tabular}
\end{center}

The two chunks run at different source-log moments. No single physical
snapshot of the table has been taken. Chunk A saw Bob at \$100. By the
time chunk B was reading, Bob's balance had already moved to \$120.

Although both chunk reads observe live rows here, a chunk read can
also record a key as \emph{absent}. If chunk B had executed before Carol's
$c_2$ insert, it would have observed account 300 as absent,
and that recorded absence is replayed identically to a concrete value.

\subsection{The Chosen Point of
Equivalence}\label{the-chosen-point-of-equivalence}

The backfill targets a specific source-log moment for the assembled
replay to match. We designate
\textbf{frontier $f = c_4$}, representing the source-log position
immediately after the last write in the sequence. The \textbf{scope} of the
backfill comprises the claimed accounts, $K = \{100, 200, 300\}$
(Alice, Bob, Carol).

Deterministic replay of the gathered events equals the source state at
$f$ on $K$, although no chunk is read at $f$, the three accounts are
never read together, and source writes continue throughout.

\subsection{How the Model Assembles the
Replay}\label{how-dblog-assembles-the-replay}

The model treats every observed source-log event as an ordinary
\textbf{CDC event} (the captured record of one committed source
write, such as an insert, update, or delete at its source-log coordinate),
every chunk-read row as a \textbf{refresh event}, and concatenates them
in source-log order. Such an event is a \emph{refresh} because it
re-states a key's current stored value, read directly from the table,
rather than reporting an incremental change as a CDC event does.

This raw replay retains every chunk refresh at its attributed read
coordinate. Section~\ref{watermarks-and-chunk-read-coordinates} explains
its relationship to DBLog's window-based suppression and output order.
For our example the assembled replay sequence is:

\begin{center}\ttfamily\scriptsize
\begin{tabular}{@{}r@{\hskip 0.4em}l@{\hskip 1em}l@{}}
1. & CDC at c1:                  & Update account 100 $\to$ 3000\\
2. & Refresh from chunk A at c1: & account 100 observed as 3000\\
3. & Refresh from chunk A at c1: & account 200 observed as 10000\\
4. & CDC at c2:                  & Insert account 300 $\to$ 2500\\
5. & CDC at c3:                  & Update account 200 $\to$ 12000\\
6. & Refresh from chunk B at c3: & account 300 observed as 2500\\
7. & CDC at c4:                  & Update account 300 $\to$ 3500\\
\end{tabular}
\end{center}

Both refreshes from chunk A sit at $c_1$. The refresh from
chunk B sits at $c_3$. CDC events sit at the source-log
coordinate they were committed at. Where a CDC event and a chunk
refresh share a coordinate (as item~1 and items~2--3 all sit at
$c_1$), the CDC event is placed first, because a chunk read at a
coordinate already reflects every write committed at that coordinate.
Figure~\ref{fig:running-example-timeline}
places these chunk reads and CDC events on the source-log timeline.

This is the \textbf{raw replay} form. It shows the chunk refreshes DBLog
gathered even when a later CDC event for the same key supersedes them.
The normalized clean prefix may omit refreshes dominated by later CDC
events, such as item~3 for account 200 and item~6 for account 300. Applying
either the raw or the normalized form yields the same per-key state. The raw sequence
shows what DBLog observed, while the normalized sequence makes stale-refresh
suppression explicit for the formal model.

To reconstruct per-account state, we play the seven raw items left to
right, each one writing the per-account map. When two items concern the
same account, the later one wins for that account's value.

Tracing the replay sequence in order:

\begin{itemize}
\tightlist
\item
  Item 1 sets account 100 to 3000.
\item
  Item 2, chunk A's refresh of account 100, confirms 100 = 3000. No
  change.
\item
  Item 3, chunk A's refresh of account 200, sets account 200 to 10000.
\item
  Item 4 introduces account 300 at 2500.
\item
  Item 5 sets account 200 to 12000. This overrides chunk A's earlier
  observation: chunk A observed balance \$100 at $c_1$, which the later
  log update at $c_3$ (\$120) supersedes.
\item
  Item 6, chunk B's refresh of account 300, confirms 300 = 2500. No
  change yet.
\item
  Item 7 sets account 300 to 3500. This overrides chunk B's earlier
  observation: chunk B observed \$25 at $c_3$, which the update at
  $c_4$ (\$35) likewise overrides.
\end{itemize}

The final per-account state, restricted to scope $K = \{100, 200, 300\}$, is

\begin{center}\ttfamily\footnotesize
\begin{tabular}{@{}l@{\hskip 1em}l@{}}
account 100 & balance = \phantom{0}3000\\
account 200 & balance = 12000\\
account 300 & balance = \phantom{0}3500\\
\end{tabular}
\end{center}

The source state at $f = c_4$ restricted to $K$ is

\begin{center}\ttfamily\footnotesize
\begin{tabular}{@{}l@{\hskip 1em}l@{}}
account 100 & balance = \phantom{0}3000\\
account 200 & balance = 12000\\
account 300 & balance = \phantom{0}3500\\
\end{tabular}
\end{center}

They match. The assembled replay reproduces the source state at $f$
across all accounts in scope $K$, even though chunk A and chunk B were read
at distinct earlier coordinates, and even though both chunks contained rows
that had become stale before reaching $f$.

\subsection{Advancing the Cut to a Later
Frontier}\label{sec:advancing-cut-running-example}

The seven-item replay above forms a virtual cut at frontier $f = c_4$
on scope $K = \{100, 200, 300\}$, matching the source state at that
point. Backfill operations are not restricted to this initial cut.
Once all chunks in the plan have been processed (here, chunks A and B),
chunk scanning terminates.
However, as the source continues committing writes, DBLog keeps consuming
the source log, allowing the cut to advance.

Suppose another write commits after $c_4$:

\begin{center}\footnotesize
\begin{tabular}{@{}l@{\hskip 1em}l@{\hskip 1em}l@{}}
\ttfamily c5 & \ttfamily Delete 200 & \itshape (Bob closes the account)\\
\end{tabular}
\end{center}

DBLog does not re-read a chunk to account for this change. The single
CDC event committed after $c_4$ and through $c_5$ is the
\emph{continuation segment} for that interval. Appending that segment
advances the cut, provided that the CDC feed delivers it accurately and
the source log still retains the interval. Section~\ref{source-side-continuation-and-restriction}
states these conditions precisely.
The replay sequence gains one event:

\begin{center}\ttfamily\footnotesize
\begin{tabular}{@{}r@{\hskip 0.4em}l@{\hskip 1em}l@{}}
8. & CDC at c5: & Delete account 200\\
\end{tabular}
\end{center}

Playing item 8 after the seven-item replay makes account 200 absent. As
with an insert or update, the later source-log event overrides both the
earlier refresh and the $c_3$ update. For a delete, the resulting state
is absence.

The eight-item replay is now a virtual cut at the later frontier
$f' = c_5$. Its per-account state on $K$ is account 100 = 3000, account
200 absent, account 300 = 3500, which is the source state at $c_5$ on
$K$. The point of equivalence moved from $c_4$ to $c_5$ by appending the
one event the source committed in between. No chunk was re-read. Had
that event been an ordinary insert or update rather than a delete, it
would have appended in exactly the same way. This is
the running example's instance of the continuation result of
Section~\ref{source-side-continuation-and-restriction}. Ongoing DBLog
execution thus produces an evolving sequence of virtual cuts, where each
successive cut advances the prior state by appending committed source
changes.

This says nothing, by itself, about whether a destination consuming the
replay has applied item 8 or reached the state in which account 200 is
absent. That remains a separate sink-side question
(Section~\ref{sink-state-convergence-non-claim}).

\begin{figure*}[t]
  \input{figures/fig02_running_example_timeline/fig02.tex}
\end{figure*}

\subsection{What ``Snapshot-Equivalent'' Means}\label{what-snapshot-equivalent-means-here}

At no point in this example did DBLog read the entire table at one
source-log position. Chunk A saw account 100 and account 200 at
$c_1$. Chunk B saw account 300 at $c_3$. Because the two reads
happened at different points in the source database's commit order,
DBLog must reconcile them through the source log rather than read them
as one instantaneous observation.

Instead, DBLog captures sufficient source-log evidence around each
chunk read to determine which observations remain authoritative at the
chosen frontier and which have been superseded. The assembled replay
sequence combines chunk refreshes and CDC events deterministically to
yield the exact per-key state of the source at frontier $f$ on scope $K$.

In this paper, \emph{snapshot-equivalent replay} means that the replay
outcome equals the source state at $f$ restricted to $K$. It does not
mean that DBLog performed a physical snapshot read.

\subsection{Scope of the Example}\label{what-this-example-does-not-show}

The example covers one three-key backfill. It establishes replay
equality for that scope, but it does not extend to the following properties:

\begin{itemize}
\tightlist
\item
  \textbf{It is not a physical cut.} DBLog never performed a single read
  covering all of \{100, 200, 300\}. Chunk A read accounts 100 and 200 at
  $c_1$, and chunk B read account 300 at $c_3$. The equivalence is
  \emph{extensional}. Outcomes match at $f$ on $K$, although DBLog did
  not take a snapshot.
\item
  \textbf{It is not a whole-table claim.} The example claims only the
  three accounts in $K$. Other accounts in the source table are not
  addressed by this example. Whole-table correctness is a derived
  specialization handled later in the paper, with extra conditions
  attached.
\item
  \textbf{It is not a sink-state claim.} What we have shown is the
  assembled replay sequence. Whether a sink that consumes
  this sequence converges to the same per-account state is an additional
  contract on the consumer, including idempotency, ordering, and
  fencing. The example does not establish that contract.
\end{itemize}

\subsection{Required Coordination}\label{what-coordination-dblog-still-needs}

Avoiding a physical snapshot read does not remove the need for
coordination. The example relies on four source-side conditions:

\begin{itemize}
\tightlist
\item
  A \textbf{source-log ordering} that is monotone and well-defined
  enough to compare the chunk-read coordinate of chunk A
  ($c_1$) to the CDC event for account 200 at
  $c_3$ and conclude that $c_3$ is later.
  Without a comparable ordering on commit positions, the ``later wins''
  rule cannot be applied.
\item
  \textbf{Chunk-read watermarks} placed in the source log around each
  chunk read, allowing DBLog to locate chunk A's read in source-log order.
  The watermark bracket places its observation of account 200 before the $c_3$
  event for account 200, identifying the event as strictly newer.
  The watermark bracket bounds the read's position in source-log order,
  as described in the next section.
\item
  \textbf{Source-log retention.} The source log must retain each
  required event until DBLog has captured it. If the $c_3$ update to
  account 200 were lost before it was captured, Bob's stale chunk value
  would survive and the replay would not match the source at $f$.
\item
  \textbf{Chunk-read consistency.} Chunk A's two rows must both reflect
  the source at the single coordinate $c_1$. The watermark bracket
  locates the read in source-log order but does not by itself
  establish that single-coordinate attribution. The deployment must
  establish this property, through a consistent statement-level read or
  through per-row evidence that all returned row images match one
  common source coordinate.
  Section~\ref{watermarks-and-chunk-read-coordinates} makes this
  obligation precise.
\end{itemize}

The deployment must establish these conditions. The formal model
records them as explicit predicates and assumptions.

\subsection{Connection to the Formal Model}\label{what-comes-next}

Section~3 defines the operational terms used above, including chunks,
watermarks, refresh events, CDC events, replay prefixes, frontiers, and
scopes. Sections~5 through~11 then give these terms formal definitions
and prove the general results illustrated here. The seven-item replay
matches the source at $f=c_4$ on $K$. Appending the eighth item advances
the same equality to $f'=c_5$.

\section{The DBLog Mechanism}\label{the-dblog-mechanism}

The running example used two chunks and four source-log events. We now
define the operational parts of the general mechanism. Figure~\ref{fig:dblog-mechanism}
shows the example on a source-log axis, including the chunk reads,
refresh windows, and merged raw replay.

\begin{figure*}[t]
  \input{figures/fig03_dblog_mechanism/fig03.tex}
\end{figure*}

\subsection{Chunks and Scope}\label{chunks-and-scope}

DBLog begins with a \textbf{scope}, which is the set of primary keys for
which the backfill will make a claim. In the example, the scope was the three
account ids \{100, 200, 300\}. In a larger deployment the scope might be
a table, a shard, a tenant, or a selected key range.

The scope is partitioned into \textbf{chunks}. The chunks are pairwise
disjoint and together cover the claimed scope. Each key in the scope
therefore belongs to exactly one chunk. The deployment chooses the
chunk ranges, and the model takes that plan as an input. A chunk may
also cover keys that do not yet exist in the table. An observed key absence
within a chunk range is recorded as evidence just like an existing row.

Unique ownership is required by the current model. If two chunks own
the same key, or if no chunk owns a key in scope, the replay lacks a
uniquely defined refresh source for that key. Overlapping and retried chunks
require a normalization step and are left as a future extension.

The model also abstracts away how the finite chunk plan is discovered.
A deployment must still know when the chunk loop has reached the
intended end of the claimed key scope. For a whole-table run, a common
way to establish that fact is to record the table's maximum primary key,
or an equivalent end-of-table observation, with enough source-side
context to know which table population the bound refers to. For a
selected range, shard, tenant, or other scoped run, the analogous
requirement is an end-of-scope marker for that chosen scope. The
deployment must provide this evidence. The model separately assumes that
the primary-key order and the source observation used for the bound are
accurate.

\subsection{Watermarks and Chunk-Read
Coordinates}\label{watermarks-and-chunk-read-coordinates}

\textbf{Watermarks} bound the source-log interval containing a chunk read.
DBLog uses this interval without needing to identify the read's exact
position within it. Before a chunk
read, DBLog writes a \emph{low watermark} to a dedicated table in the
source. After the scan, it writes a \emph{high watermark}. Both writes
appear in the source log at definite coordinates, with the chunk
\texttt{SELECT} between them. The baseline mechanism therefore needs
write access to the source. A read-only deployment instead relies on
existing source-log coordinates, such as transaction identifiers or LSN markers.

The watermark pair establishes the read's \emph{provenance} in
source-log order. The formal model attributes the returned rows to one
\emph{chunk-read coordinate}, such as $c_1$ for chunk A in the running
example. Every row image is treated as matching the source at that
coordinate.

The watermark bracket alone does not establish this single-coordinate
attribution. A multi-row \texttt{SELECT} is not instantaneous. The
deployment must also provide either read semantics that make all rows
consistent at one source coordinate, such as a statement-level snapshot
or repeatable-read scan, or per-row evidence that all returned row images
match one common source coordinate. Thus, ``chunk A
read at $c_1$'' combines two distinct conditions: the watermarks
bracket the read around $c_1$, and the returned row images accurately
reflect the source state at $c_1$.

In the original DBLog mechanism, the interval between the two
watermarks is also the window over which chunk rows with matching CDC
events are suppressed. A chunk row is discarded when the source log shows a write
for the same key inside that window, whether that write committed
before or after the chunk read. The model represents that behavior with
two simpler parts. Each returned row has a valid chunk-read coordinate,
and replay applies a frontier-wide \emph{latest-event-wins} rule
(\S\ref{refresh-events-and-ordinary-cdc-events}). The raw replay keeps
every chunk row.

For a completed watermark window whose CDC events have been captured
in source order, both formulations reach the same per-key state at the
window close. For a key in the chunk's domain, there are three cases. If a
same-key write commits after the chunk read but within the window, the
last such write determines the key's final state in both formulations. If the
only same-key writes inside the window commit before or at the chunk
read, the read already reflects them. The chunk row then carries the
same value or absence as the last of those writes, and discarding it
changes nothing. If no same-key write falls inside the window, the row
survives in both formulations. Original DBLog emits it after the high
watermark, while the model places it at the chunk-read coordinate. No
write to that key lies between the two positions, so the per-key result
is the same. A write after the high watermark overrides the row in both
formulations.

The Isabelle development mechanizes the first case as the
dominated-refresh equivalence
(\S\ref{raw-replay-and-normalized-clean-prefix}). The other two cases use refresh
fidelity, a wellformed-run obligation, together with source-state and
$\Apply$ lemmas the development proves. It does not state the
window-discard transformation as a separate theorem.

The formal run retains the low and high watermarks as evidence for the
chunk-read coordinate. They do not add another state-equality
condition. Section~\ref{wellformed-runs} records the required source-log
ordering, retention, coordinate provenance, and row-image accuracy.
The last condition appears there as ``Refresh fidelity.''

DBLog avoids a single physical snapshot read. It does not avoid
source-log coordination.

\subsection{Refresh Events and Ordinary CDC
Events}\label{refresh-events-and-ordinary-cdc-events}

The model represents each chunk-read row as a \textbf{refresh event}. A refresh
records that key $k$ held a given value, or was absent, at the
chunk-read coordinate. A source-log write becomes an ordinary CDC event
that records an insert, update, or delete at its source-log coordinate.

Both kinds of event enter one replay vocabulary. Ordinary CDC events
carry committed source writes. Refresh events carry row observations
from chunk reads. Applying a replay means walking the event sequence
left to right and updating the per-key state. Later events for the same
key win. That behavior is what made Bob's $c_3$ update
override chunk A's older refresh in the example.

\subsection{Raw Replay and Normalized Clean
Prefix}\label{raw-replay-and-normalized-clean-prefix}

The model constructs a finite sequence called the \emph{clean prefix}. We use
two forms of this sequence. The theorems use the \emph{raw} form. The
\emph{normalized} form is an optional, $\Apply$-equivalent presentation.

The \textbf{raw replay} is the evidence-preserving sequence. It may
include a refresh even if a later CDC event for the same key dominates
it. Raw replay is useful operationally because it shows exactly what
DBLog observed.

The \textbf{normalized clean prefix} makes stale-refresh suppression
explicit. It may omit refresh events that are dominated by newer
source-log evidence, provided suppressing them does not change the final
$\Apply$ state. In the example, Bob's chunk-A refresh was
dominated by Bob's later c3 update. Carol's chunk-B refresh was
dominated by Carol's later c4 update. Carrying or suppressing those
dominated refreshes gives the same final per-key state. The underlying
formal connection is state equivalence over the claimed scope. Suppressing
dominated refreshes for the same key preserves the replay outcome under
$\Apply$ for all claimed keys.

\smallskip
\noindent\textbf{Dominated-refresh equivalence.} \emph{A refresh
dominated by a same-key CDC event at a strictly later source coordinate
may be retained or suppressed without changing $\Apply$ on any claimed
key.} This fact connects the raw and normalized prefixes. It also covers
the later-write case of original DBLog's in-window suppression. The
case of a write before or at the chunk read is argued separately in
\S\ref{watermarks-and-chunk-read-coordinates}.
Section~\ref{source-side-virtual-cut-soundness-layer-2} states the
Layer~2 theorem about the raw replay prefix and uses this equivalence
to extend its conclusion to the optional normalized presentation.

The raw replay and its optional normalization satisfy three properties:

\begin{itemize}\tightlist
\item no required CDC event is missing.
\item no stale refresh is allowed to resurrect an older value past a
  newer event.
\item suppressing a refresh is justified only when newer source-log
  evidence dominates it for the same key.
\end{itemize}

\subsection{Frontier and Clean Prefix}\label{frontier-and-clean-prefix}

The \textbf{frontier} is the source-log coordinate at which a cut is
claimed (concretely, a MySQL executed global-transaction-identifier
(GTID) set, a PostgreSQL log sequence number (LSN), or a
CockroachDB resolved timestamp). A certificate claims equality at a
particular frontier, not at an unspecified eventual state. The clean prefix contains the
replay evidence needed through that frontier for the claimed scope.

Operationally, a DBLog certificate asserts that for a given scope $K$
and frontier $f$, replaying the certified prefix reconstructs the exact
source state at $f$ on $K$.

The rest of the paper makes that sentence precise. Section~5 defines the
source and replay models. Section~6 names the equality as a virtual cut.
Section~7 proves the source-side theorem for wellformed runs. Sections~8
and~9 move from runs to accepted certificates.

The clean prefix is finite but not necessarily small. It contains a
refresh for each key in the claimed scope and every in-scope CDC event
through the frontier, so it scales with the scope and the change
volume over the certified interval. For a large table backfilled
over a busy window, the clean prefix is correspondingly large. The
normalized clean prefix trims refreshes that newer evidence dominates.
The raw form, on which the theorems are stated, does not.

\subsection{Advancing the Frontier Beyond the Initial
Backfill}\label{after-the-chunk-loop}

Once the chunk loop covers the claimed scope and terminates, DBLog
continues consuming ordinary CDC events as the source commits them. The
source-side result does not need a fresh chunk read for every later
frontier. If the backfill has established a virtual cut at $f$, appending
exactly the in-scope source events committed after $f$ and through a later
frontier $f'$, in source order, advances that cut to $f'$.

The continuation theorem formalizes this step in
Section~\ref{source-side-continuation-and-restriction}. Once the backfill
establishes an initial virtual cut, subsequent CDC stream events advance
it. The chunk reads happen once. Later cuts are derived by appending
valid source-log change events. This step relies on ordered source-log
coordinates and retention of the events between $f$ and $f'$.
Section~\ref{source-side-continuation-and-restriction} states the result
and its assumptions.

DBLog combines a finite chunk plan, source-log watermarks,
chunk refreshes, and ordinary CDC events into an equivalent replay at a
named frontier. The deployment must provide an ordered and retained
source log, accurate chunk reads, and complete coverage of the claimed
scope.

\section{DBLog Claims and
Limitations}\label{what-dblog-is-what-it-is-not-and-its-limitations}

DBLog combines source-log events with chunk-read observations. A live
backfill establishes an initial virtual cut, and later CDC events can
advance that cut to new source frontiers. The result is a source-side
replay claim on a named scope and frontier. It does not remove the need
for source coordination or establish the state of a destination.

\subsection{What DBLog Provides}\label{what-dblog-is}

The protocol combines six primary inputs:

\begin{itemize}\tightlist
\item a claimed key scope.
\item a certified frontier in source-log order.
\item chunk reads whose observations become refresh events.
\item ordinary CDC events from the source log.
\item source-log and chunk-read evidence sufficient to justify the
  clean prefix.
\item a finite replay prefix whose $\Apply$ state can be compared with
  the source.
\end{itemize}

Replaying the clean prefix yields the exact per-key state of the source
at the frontier on the claimed scope. Once this equality has been
established, appending the source changes committed after that frontier
and through a later frontier advances the cut. Neither step requires a
physical snapshot read.

\subsection{Limits of the Result}\label{what-dblog-is-not}

The result depends on source-log ordering, retained CDC events,
chunk-read coordinates, frontier ordering, and evidence that connects
these observations to the source. A deployment or source connector must
establish these conditions.

The core theorem guarantees equality specifically for the claimed scope.
Whole-table equality is a derived result requiring the Layer~4
anchor-domain and no-unclaimed-replay-key conditions.

The theorem does not establish destination state or exactly-once delivery.
A destination may receive the certified replay without applying
it in the required order or reaching the corresponding state. The
source-side continuation result advances replay equality to later
frontiers, but it does not show that a sink has applied those events.
Section~\ref{sink-state-convergence-non-claim} states this limit.

A virtual cut also does not imply that one physical source-read
timestamp existed for all chunk rows. It is an equality of outcomes.

Debezium, Apache Flink CDC, MySQL, PostgreSQL, and MariaDB provide
deployment and related-work context. The theorem does not apply to any
named system until its implementation and source assumptions have been
checked against the formal model.

\subsection{Failure Modes and Required
Evidence}\label{sec:failure-modes-evidence}

Table~\ref{tab:failure-modes} lists the main failure modes, the evidence
or assumption that addresses each one, and whether the condition is
machine-checked or supplied by the deployment.

\begin{table*}[t]
\caption{Failure modes, required evidence or assumptions, and their
status.}
\label{tab:failure-modes}
\centering\small
\begin{tabularx}{\textwidth}{@{}p{1.55in}YYY@{}}
\toprule
\textbf{Failure mode} & \textbf{What goes wrong} &
\textbf{Evidence or assumption boundary} & \textbf{Status}\\
\midrule
Missing CDC event &
  A source write in the claimed interval is absent from the replay
  evidence. &
  Source-log retention and CDC coverage through the certified
  frontier. &
  Deployment obligation / External observation assumption\\
Missing continuation segment &
  A later frontier is claimed without the actual source-log change
  data the source committed between the certified frontier and that
  later frontier. &
  Source-log retention for the continuation interval and complete CDC
  coverage. &
  Deployment obligation / External observation assumption\\
Bad watermark or coordinate &
  A chunk row is treated as if it was read at the wrong source-log
  position. &
  Watermark or coordinate validity for the chunk read. &
  Deployment obligation / External observation assumption\\
Stale refresh survives &
  An older chunk-read value overrides a newer source-log event for
  the same key. &
  Clean-prefix dominance and latest-event ordering. &
  Machine-checked\\
Wrong frontier &
  The replay is certified against a frontier different from the
  source interval it actually covers. &
  Frontier consistency in the certificate and evidence. &
  Machine-checked\\
Scope ownership mismatch &
  A claimed key is not owned by exactly one responsible chunk, or the
  claimed scope does not match the chunk plan. &
  Chunk ownership and claim-scope checks. &
  Machine-checked\\
Incomplete chunk-plan termination &
  The chunk loop stops before the intended end of the key range, so a
  key in the intended operational scope may be absent from the finite
  plan. For a whole-table claim, the anchor domain is not shown to be
  covered. &
  Chunk-plan construction, maximum-key or end-of-scope evidence,
  primary-key ordering, and whole-table claim-scope coverage. &
  Deployment obligation / External observation assumption\\
Whole-table replay-key mismatch &
  The whole-table specialization is requested while replay evidence
  materializes keys outside the claim scope. &
  Layer 4 no-unclaimed-replay-key side condition, which is not part of scoped
  virtual-cut equality. &
  Machine-checked\\
Inaccurate observation &
  The connector, source log, primary-key interpretation, or
  chunk-read result does not accurately reflect the source database. &
  Not internally provable from the certificate alone. &
  External observation assumption\\
\bottomrule
\end{tabularx}
\end{table*}

For the initial certificate, a checker can expose the internal evidence
failures listed in the table. Layer~4 adds the conditions needed for a
whole-table claim. A concrete checker for continuation certificates is
not included. In all cases, the theorem assumes that the source
observations match the mathematical source history.

\subsection{Trust Boundary}\label{sec:trust-boundary}

The strongest result remains source-side. Under accurate source
observation and the checker assumptions, accepted evidence yields a
certified virtual cut. The result does not prove that a concrete parser
is correct, that a connector reports the source accurately, that a
named product implements every assumption, or that a sink has applied
the replay.

\section{Formal Model \mbox{(Layer 0 and Layer 1)}}\label{formal-model-layer-0-and-layer-1}

Layer~0 defines source histories, source state at a frontier, replay
events, replay application, and restriction to a key scope. Layer~1
defines abstract DBLog runs, including chunks, watermarks, chunk reads,
CDC observations, clean prefixes, and wellformedness conditions.

These layers do not include certificates, verifiers, destination ACKs,
or whole-table conditions. Layer~2 proves a theorem about wellformed
runs. Layer~3 then establishes when an accepted certificate guarantees
such a run.

Table~\ref{tab:notation} fixes the symbols the formal sections use. A
predicate (short sans-serif tag like $\WF$, $\FSO$, $\Mat$, $\Coh$,
$\VC$, $\WTS$) is a relation that holds or fails on its arguments. An
accessor function (italic word like $\Apply$, $\Src$, $\Scope$,
$\Frontier$, $\CleanPrefix$) returns a value.

\begin{table*}[t]
\caption{Notation used in the formal model and theorem statements.
Predicates are rendered in sans-serif. Accessor functions are italic.}
\label{tab:notation}
\centering\small
\begin{tabularx}{\textwidth}{@{}l@{\hskip 1em}l@{\hskip 1em}Y@{}}
\toprule
\textbf{Symbol} & \textbf{Kind} & \textbf{Read as}\\
\midrule
$b_0$, $H$, $f$ & variables &
  base state, source history, frontier coordinate.\\
$b$, $f'$ & variables &
  whole-table anchor source coordinate ($b \le f$) and later source
  frontier as the continuation target ($f \le f'$).\\
$R$, $C$, $E$ & variables &
  abstract DBLog run, certificate, evidence bundle (\S\ref{evidence}).\\
$\sigma$, $\delta$, $K$ & variables &
  replay sequence, continuation segment (CDC replay-event list),
  key scope.\\
$\Src(b_0,H,f)$ & accessor & source state at frontier $f$.\\
$\Apply(\sigma)$ & accessor & replay-apply of sequence $\sigma$.\\
$\Scope(\cdot)$ & accessor & claimed key scope (of run or certificate).\\
$\Frontier(\cdot)$ & accessor & chosen source-log frontier.\\
$\CleanPrefix(\cdot)$ & accessor & clean replay prefix.\\
$m \restrict K$ & operator & state $m$ restricted to keys in $K$.\\
$\Verify(C,E)$ & accessor &
  verifier result, one of $\Accept$, $\Reject$, $\Unsupported$.\\
$\RawObs$ & variable &
  raw source-observation carrier (External observation assumption).\\
$\AnchorDomain(H,b)$ & accessor & keys present at the anchor boundary $b$.\\
$\TouchedBetween(H,b,f)$ & accessor & keys with source events in $(b,f]$.\\
$\TableScopeBetween(H,b,f)$ & accessor & union of the two above.\\
$\ReplayKeys(\sigma)$ & accessor & keys written by replay sequence $\sigma$.\\
$\Chunks(R)$ & accessor & run's chunk plan.\\
$\SourceHistory(R)$ & accessor & source history the run is about.\\
$\CdcEvents(R)$ & accessor & CDC events the run consumed.\\
\midrule
$\WF(b_0,R,H)$ & predicate &
  wellformed DBLog run (Layer 1).\\
$\VCS(\sigma,K,f,H)$ & predicate &
  state-level virtual cut (Section~\ref{virtual-cuts}). The base state
  $b_0$ is implicit in this four-argument shorthand and made explicit
  in the unfolded form in Section~\ref{virtual-cuts}.\\
$\VC(C,H)$ & predicate &
  certificate-side virtual cut.\\
$\ContSeg(H,K,f,f',\delta)$ & predicate &
  $\delta$ is the continuation segment for $(f,f']$ on $K$
  (Section~\ref{source-side-continuation-and-restriction}).\\
$\Mat(C,E,R)$ & predicate &
  $(C,E)$ materializes abstract run $R$.\\
$\Coh(C,E,R)$ & predicate &
  $(C,E)$ is coherent with run $R$ (Section~\ref{certificate-run-coherence}).\\
$\FSO(E,b_0,H,\RawObs)$ & assumption &
  accurate source observation ($\FSO$).\\
$\WTS(C,H,b)$ & predicate &
  whole-table claim scope (Layer 4).\\
\bottomrule
\end{tabularx}
\end{table*}

\subsection{Source Histories and
Frontiers}\label{source-histories-and-frontiers}

A source database is modeled as a base state $b_0$ together with a
finite history $H$ of committed source events. A source event is
one of
\[
  \mathsf{Insert}\,k\,v,\qquad
  \mathsf{Update}\,k\,v,\qquad
  \mathsf{Delete}\,k.
\]
Each event records a source coordinate, such as a commit position, log sequence
number, monotonic timestamp, or equivalent order-bearing
token. We write $c \le c'$ when
coordinate $c$ is no later than $c'$. A
source history is wellformed when its events are ordered by that
coordinate, the base coordinate sits before every event, and
same-coordinate events have a deterministic source-position order inside
the history.

The source state at a frontier is written $\Src(b_0,H,f)$.
For each key $k$, this state looks for the latest source event
for $k$ at or before frontier $f$. An insert or update
makes the key present with the written value. A delete makes it absent.
If there is no event, the key keeps its base state $b_0(k)$.

The coordinate convention is inclusive. A chunk read recorded at
coordinate $c$ observes $\Src(b_0,H,c)$, including
source events at coordinate $c$. A chunk read is \emph{attributed} to
a coordinate within its watermark bracket at which all its row images
match the committed source state.
\S\ref{watermarks-and-chunk-read-coordinates} sets out what a deployment
must supply for that attribution to be sound. The watermark bracket
locates the read, while chunk-read semantics or coordinate evidence
establish that all returned row images match that common
coordinate. Canonical replay therefore
places same-coordinate CDC evidence before the refresh that records the
post-$c$ chunk-read state. Later source-coordinate events still
appear after that refresh.

This is the source-side reference state that every virtual-cut theorem
compares against.

\subsection{Replay Events and Apply}\label{replay-events-and-apply}

DBLog does not replay source events alone. It replays two kinds of
replay event:
\[
  \begin{array}{ll}
    \mathsf{Cdc}\,c\,e & \text{source-log event $e$ at coord.\ $c$,}\\
    \mathsf{Refresh}\,k\,m\,c & \text{chunk-read of $k$ at coord.\ $c$.}
  \end{array}
\]
In a refresh, $m = \mathsf{Some}\,v$ means the chunk read observed
value~$v$. The value $m = \None$ means that the row was absent.
Absence is part of the state. Replaying a refresh
with $\None$ deletes the key from the replay state.

The replay function is $\Apply(\sigma)$, where $\sigma$ is a finite
replay sequence. $\Apply$ walks the sequence from left to right,
updating a per-key map. Later events
for the same key win. $\Apply$ starts from the all-absent per-key
map. Formally, $\Apply([\,])(k) = \None$ for every key~$k$. Thus the running
example's later update for Bob overrides chunk A's older refresh, and
a later delete would override both an earlier refresh and an earlier
update.

State restriction is written $m \restrict K$ in prose formulas
below. It keeps only the entries whose keys lie in $K$, so
$(m \restrict K)(k) = m(k)$ when $k \in K$, and
$\None$ otherwise. The displayed formulas use the
$\restrict$ symbol throughout.

\begin{table*}[t]
\caption{Wellformed-DBLog-run obligations in $\WF(b_0,R,H)$ and their status.}
\label{tab:wellformed-obligations}
\centering\small
\begin{tabularx}{\textwidth}{@{}p{1.45in}Yp{1.8in}@{}}
\toprule
\textbf{Obligation} & \textbf{Meaning} & \textbf{Status}\\
\midrule
Source binding &
  $R$ is about source history $H$, and $H$ is wellformed. &
  Deployment obligation\\
Chunk partition &
  The chunks form a finite strict partition of $\Scope(R)$. &
  Deployment obligation\\
CDC coverage &
  Every relevant source event after the responsible chunk read and
  through the frontier is observed. &
  External observation assumption\\
CDC fidelity &
  Observed CDC events are real source events, in the right order, with
  same-key same-coordinate multiplicity preserved. &
  External observation assumption\\
Chunk evidence &
  Chunk read coordinates are bracketed by watermarks, and chunk-read
  observations produce the corresponding value-or-absence refreshes for
  claimed keys. &
  Deployment obligation\\
Read-before-frontier &
  Every chunk read occurs no later than the run frontier. &
  Deployment obligation\\
Refresh fidelity &
  A chunk-read value or absence matches
  $\Src(b_0,H,\text{read-coord.})$ for that key. &
  External observation assumption / Deployment obligation\\
\bottomrule
\end{tabularx}
\end{table*}

\subsection{Abstract DBLog Runs}\label{abstract-dblog-runs}

A DBLog run $R$ is modeled abstractly, characterized through six accessor functions:
\[
  \begin{aligned}
    \Scope(R)\quad             &\text{claimed key scope},\\
    \Frontier(R)\quad           &\text{chosen source frontier},\\
    \CleanPrefix(R)\quad        &\text{replay sequence of the run},\\
    \Chunks(R)\quad             &\text{run's chunk plan},\\
    \SourceHistory(R)           &\ \text{source history the run is about},\\
    \CdcEvents(R)               &\ \text{observed CDC events consumed.}
  \end{aligned}
\]
Each chunk has a domain of keys, a responsible-chunk relation from
keys to chunks, a lower and upper watermark, and a chunk-read
coordinate. The model uses a strict canonical partition. Each key in
$\Scope(R)$ belongs to exactly one responsible chunk, and every
responsible chunk has a finite domain. The model also takes the chunk
plan itself to be finite, so every run has a finite set of chunks. This
models a backfill with a bounded chunk loop over a finite plan.

That strict partition is not a claim that real systems never retry or
overlap chunk reads. It is the canonical run selected for the theorem.
Raw overlap/retry evidence must be normalized into such a canonical run,
or handled by a future extension.

\subsection{Clean Prefix}\label{clean-prefix}

The clean prefix of a run is the finite replay sequence constructed from three elements:

\begin{itemize}\tightlist
\item chunk-read refreshes for the responsible chunks.
\item observed CDC events for in-scope keys through the frontier.
\item the \emph{clean-prefix ordering}, sorting events by source
  coordinate, placing CDC events before chunk refreshes at
  shared coordinates, and keeping same-coordinate CDC events in
  source-log position order.
\end{itemize}

The clean prefix keeps every chunk refresh. The optional normalized
presentation drops dominated refreshes without changing the final
replay state.

The theorem does not rely on an unstated ``good replay'' primitive. The
clean prefix is constrained by the run's chunk plan, CDC observations,
frontier, and scope. If a needed CDC event is missing, if a future event
is replayed, if a refresh appears without a responsible chunk
observation, or if same-coordinate CDC events are reordered, the run is
not wellformed for the theorem.

The main invariant is per-key. For each key in scope, the
latest replay event in the clean prefix is either the latest accurate
source event through the frontier, or it is a refresh whose value or
absence at its read coordinate remains unchanged until the frontier.
Layer~2 derives its result from this invariant.

The mechanization has one additional assumption. The canonical clean
prefix enumerates each chunk domain in sorted key
order, so the run- and certificate-layer theorems assume a linear
order on the key type. This introduces no additional requirement, as chunk
selection already assumes linearly ordered primary keys. The state-level continuation and
restriction theorems of
Section~\ref{source-side-continuation-and-restriction} carry no such
constraint.

\subsection{Wellformed Runs}\label{wellformed-runs}

We write $\WF(b_0,R,H)$ for the wellformed DBLog run predicate. It
collects the source-side conditions used by the proof.
Table~\ref{tab:wellformed-obligations} lists these conditions.

The split between Deployment obligations and External observation
assumptions is intentional. The model defines what a wellformed run
must satisfy. It cannot prove, from within the run model itself, that an external
database connector reported every source event without loss or that a chunk
read accurately captured the real source state.

Refresh fidelity includes both labels because it combines two aspects. A
returned row image must report the source value or absence accurately,
which is an External observation assumption. The chunk-read semantics
pin those images to a single coordinate
(\S\ref{watermarks-and-chunk-read-coordinates}), which is a Deployment
obligation.

The predicate has one further conjunct, clean-prefix closure, omitted
from the table because it follows from the construction in
Section~\ref{clean-prefix} rather than from a deployment or external
observation. The proof may therefore use it without adding a deployment
obligation.

The lower and upper watermarks introduced above are source-log
coordinates, not primary-key-range bounds. Accordingly, the Chunk
partition row is a statement about the finite canonical partition
consumed by the theorem. It is not a proof that a concrete
chunk-selection loop stopped in the right place, and Layer 1 contains no
maximum-key accessor or chunk-loop termination algorithm. A
maximum-primary-key query, an empty-next-chunk observation, a
predeclared range bound, or another end-of-scope indicator may justify the
finite partition operationally. Supplying the resulting partition,
including evidence that the loop reached the intended end of the
claimed key scope, is a Deployment obligation. The theorem consumes the
partition, not the plan-construction algorithm.

\subsection{Layer 0 and Layer 1
Lemmas}\label{layer-0-and-layer-1-lemmas}

The source and replay model establishes the elementary facts used by the
main proof:

\begin{itemize}\tightlist
\item $\Apply$ is latest-event-wins per key.
\item Restricted states are equal exactly when they agree on every key
  in the scope.
\item $\Src$ is characterized by the latest source event at or before
  the frontier.
\item A replay-side $\mathsf{Cdc}$ event has the same per-key effect as
  its source event.
\item A refresh has the source-state value at its chunk-read coordinate
  when the chunk-read fidelity assumption holds.
\item The source state at a later frontier is the source state at an
  earlier frontier updated by exactly the source events strictly
  between them. This is immediate from the preceding characterization, since
  the source events at or before $f'$ are those at or before $f$
  together with those in $(f,f']$. This interval characterization is
  what the continuation result of
  Section~\ref{source-side-continuation-and-restriction} builds on.
\end{itemize}

The run layer then contributes the bridge lemma, displayed here as a
preview of the Layer~2 theorem stated and explained in Section~7:

\[
\begin{aligned}
  \WF(b_0,R,H) \,\Longrightarrow{}&
    \Apply(\CleanPrefix(R)) \restrict \Scope(R)\\
  &{}= \Src(b_0,H,\Frontier(R)) \restrict \Scope(R).
\end{aligned}
\]

Section~6 first names the equality itself.

A wellformed run is bound to its source history. It requires a finite
chunk partition, watermark-bracketed chunk reads, and complete, accurate
CDC evidence. The following theorems apply to runs that meet these conditions.

\section{Virtual Cuts}\label{virtual-cuts}

\enlargethispage{2\baselineskip}

A virtual cut formalizes the property that replaying a prefix yields the same per-key state as the
source at a designated frontier on a given scope. It establishes an equality of
outcomes rather than requiring a simultaneous physical read cut.

We write $\VCS(\sigma,K,f,H)$ for the state-level form, with the base
state $b_0$ left implicit in this four-argument shorthand:

\[
  \VCS(\sigma,K,f,H) \,\Longleftrightarrow\,
    \Apply(\sigma) \restrict K = \Src(b_0,H,f) \restrict K.
\]

Here $\sigma$ is a replay sequence, $K$ is a key scope,
$f$ is a frontier, and $H$ is the source history. The base
state $b_0$ is fixed by context. When needed, the formulas
write it explicitly.

The certificate-side form $\VC(C,H)$ specializes the same equality to
the accessors of a certificate:

\[
\VC(C,H) \,\Longleftrightarrow\,
  \VCS\bigl(\CleanPrefix(C),\Scope(C),\Frontier(C),H\bigr).
\]

Expanding this definition:

\[
\begin{aligned}
&\Apply(\CleanPrefix(C)) \restrict \Scope(C)\\
&\qquad= \Src(b_0,H,\Frontier(C)) \restrict \Scope(C).
\end{aligned}
\]

The definition is extensional. It says that two per-key
states are equal on a scope at a frontier. It does not say that every
chunk row was read at that frontier, that one physical source snapshot
existed, that the scope is the whole table, or that any destination has
applied the replay.
Figure~\ref{fig:virtual-vs-physical-cut} contrasts a physical cut with
the virtual cut this paper proves.

\begin{figure*}[ht]
  \input{figures/fig04_virtual_vs_physical_cut/fig04.tex}
\end{figure*}

This is the exact shape illustrated by the running example. There, the
replay sequence has seven items, the scope is
$\{100,200,300\}$, the frontier is $c_4$, and
$\Apply$ produces the same balances as $\Src$ at $c_4$
on those three keys. Chunk A and chunk B were read at different
coordinates. The virtual cut does not hide that fact. It says the replay
outcome is snapshot-equivalent on the claimed keys.

The state-level form is useful because Layer 2 proves the equality for
run accessors:

\[
  \VCS\bigl(\CleanPrefix(R),\Scope(R),\Frontier(R),H\bigr).
\]

Layer 3 then proves that an accepted certificate, under accurate source
observation, certifies a run whose accessors agree with the certificate.
That is how the run-side theorem becomes the certificate-side theorem.

\section{Source-Side Virtual-Cut Soundness \mbox{(Layer 2)}}\label{source-side-virtual-cut-soundness-layer-2}

Layer 2 proves the source-side theorem:

\[
\begin{aligned}
  \WF(b_0,R,H) \,\Longrightarrow{}&
    \Apply(\CleanPrefix(R)) \restrict \Scope(R)\\
  &{}= \Src(b_0,H,\Frontier(R)) \restrict \Scope(R).
\end{aligned}
\]

\paragraph*{Theorem Summary.}
\begin{itemize}
\tightlist
\item
  \textbf{Status.} Machine-checked under the displayed
  assumptions.
\item
  \textbf{Claim.} The clean prefix of a wellformed DBLog run reaches the
  source state at the run frontier on the run scope.
\item
  \textbf{Assumptions.} The wellformed-run obligations in
  Section~\ref{wellformed-runs},
  including chunk partition, CDC coverage and fidelity, chunk-read
  fidelity, and read-before-frontier ordering.
\item
  \textbf{Boundary.} This is a source-side run theorem, distinct from
  certificate-checking, whole-table, or sink-state theorems.
\end{itemize}

Equivalently, every wellformed DBLog run satisfies the virtual-cut-state
equality at the run's frontier on the run's scope. The theorem is stated
about the run's unnormalized replay prefix. By the dominated-refresh equivalence
of \S\ref{raw-replay-and-normalized-clean-prefix}, the normalized clean
prefix yields the same $\Apply$ state and therefore the same equality.

The theorem assumes the wellformed-run predicate of
Section~\ref{wellformed-runs}, including chunk partition, frontier
bounds, CDC coverage and fidelity, and chunk-read fidelity. Each
condition is stated explicitly because omitting it allows a counterexample
(Table~\ref{tab:essential-obligations}). The predicate's remaining
conjunct, clean-prefix closure, is derived by definition. It holds
for every run by the clean-prefix construction, so the proof may use it
but no deployment need establish it separately.

\subsection{Proof Idea}\label{proof-idea}

The proof is a per-key latest-event argument.

First, restricted-state equality reduces the theorem to pointwise
equality. It is enough to show that for every key $k \in \Scope(R)$,

\[
  \Apply(\CleanPrefix(R))(k) = \Src(b_0,H,\Frontier(R))(k).
\]

Fix such a key $k$. Since the run is wellformed, $k$ has a
unique responsible chunk, and the chunk-read evidence produces a
value-or-absence refresh for $k$ that belongs to $\CleanPrefix(R)$,
including $\mathsf{Refresh}\,k\,\None\,c$ when the row is absent at the
chunk-read coordinate. Therefore the ``no replay event for $k$'' case
cannot occur for an in-scope key. By the clean-prefix construction,
every in-scope key $k$ has a latest event in $\CleanPrefix(R)$, which
must be either a CDC event or the responsible chunk's refresh. The proof
therefore has two cases.

\subsection{Case 1. The Latest Replay Event Is
CDC}\label{case-1-the-latest-replay-event-is-cdc}

Suppose the latest replay event for $k$ in
$\CleanPrefix(R)$ is a CDC event $\mathsf{Cdc}\,c\,e$. Clean-prefix
coherence ensures this replay event originates from an observed CDC pair,
while CDC fidelity guarantees that this pair corresponds to a genuine source
event in $H$. Because the event lies in the clean prefix, its
source coordinate $c$ is within the run's frontier, i.e.\ $c \le
\Frontier(R)$. Order preservation and same-coordinate
multiplicity ensure that the replay preserves both the order and the
occurrences of source events for that key and coordinate.

Now assume, for contradiction, that the source history has a later event
for $k$ at or before the frontier. CDC coverage would put that
later event into the observed CDC list. Clean-prefix closure would put
it into the replay. The clean-prefix ordering would place it after
$\mathsf{Cdc}\,c\,e$, contradicting the choice of $\mathsf{Cdc}\,c\,e$ as
the latest replay event for $k$.

So the latest replay CDC event corresponds to the latest source event
for $k$ through the frontier. $\Apply$ and $\Src$
therefore make the same update at $k$.

\subsection{Case 2. The Latest Replay Event Is a
Refresh}\label{case-2-the-latest-replay-event-is-a-refresh}

Suppose the latest replay event for $k$ is
$\mathsf{Refresh}\,k\,m\,c_{\mathit{read}}$. The responsible-chunk obligations
identify the responsible chunk, and refresh fidelity ensures that

\[
m = \Src(b_0,H,c_{\mathit{read}})(k).
\]

If there were a later source event for $k$ in
$(c_{\mathit{read}}, \Frontier(R)]$, CDC coverage and clean-prefix
closure would place a corresponding $\mathsf{Cdc}$ event after the refresh
in the clean prefix. The interval is open at $c_{\mathit{read}}$ because a
read at $c_{\mathit{read}}$ already observes source events at
$c_{\mathit{read}}$. The post-read repair obligation starts after the read
coordinate and runs through the frontier. A later source event would
contradict the refresh being latest for $k$.

Therefore the source state for $k$ at the read coordinate is
still the source state for $k$ at the frontier. The refresh
writes exactly that value or absence, so
$\Apply(\CleanPrefix(R))(k)$ equals
$\Src(b_0,H,\Frontier(R))(k)$. Both row presence and row absence
satisfy this equation, with absence represented by $\None$ in the
per-key state.

\subsection{Why the Assumptions Are
Necessary}\label{sec:assumption-necessity}

The proof fails if the obligations are weakened.

\begin{table*}[t]
\caption{Counterexamples obtained when selected obligations in $\WF$
are removed.}
\label{tab:essential-obligations}
\centering\small
\begin{tabularx}{\textwidth}{@{}p{1.4in}YY@{}}
\toprule
\textbf{Missing obligation} & \textbf{Counterexample shape} &
\textbf{Boundary protected}\\
\midrule
CDC coverage &
  A source update after a chunk read is missing from the replay, allowing a stale
  refresh value to persist. &
  DBLog needs retained source-log evidence.\\
Strict chunk ownership &
  Two chunks both claim a key, or no chunk owns it, so the refresh
  source is ambiguous or absent. &
  The theorem is scoped through a canonical chunk plan.\\
Read-before-frontier &
  A chunk read after the frontier injects a future value into the
  replay. &
  The replay is frontier-bound.\\
Same-coordinate order and multiplicity &
  Duplicate or same-coordinate source events are observed in a
  different order or with an occurrence missing. &
  Latest-event reasoning must match source order.\\
Refresh fidelity &
  A chunk refresh records a value or absence that was not the source
  state at the read coordinate. &
  Chunk reads remain an External observation assumption.\\
Clean-prefix order &
  A clean-prefix event for some key sits before an earlier-coordinate
  same-key event, so the prefix's latest event for that key differs
  from the source's. &
  Per-key latest-event reasoning transfers from the prefix to the
  source.\\
\bottomrule
\end{tabularx}
\end{table*}

These negative cases are the reason the theorem
requires the full wellformed-run predicate rather than a vague
statement that DBLog ``collected some chunks and some CDC.''

A wellformed backfill produces a replay sequence that reconstructs the
source state at the chosen frontier on the chosen scope, requiring neither long-lived table
locks nor pauses in source writes. The next two layers connect this run-level guarantee
to an inspectable certificate and, under additional conditions, extend it to whole-table
equivalence.

\section{Certificates, Evidence, and
Verifiers}\label{certificates-evidence-and-verifiers}

Layer~2 proves a theorem about an abstract DBLog run. If the run is
wellformed, its clean replay prefix reaches the source state at its
frontier on its scope. In practice, a deployment provides a certificate and associated evidence
bundle rather than an abstract run. This section defines those artifacts,
and Section~9 gives the corresponding theorem.

\subsection{Certificate}\label{certificate}

A \textbf{certificate} is the claim object, specifying three accessors:
\[
  \begin{aligned}
    \Scope(C)\quad        &\text{the keys being claimed},\\
    \Frontier(C)\quad      &\text{the source-log coordinate of the claim},\\
    \CleanPrefix(C)\quad   &\text{the certified replay prefix.}
  \end{aligned}
\]
Before acceptance, $\CleanPrefix(C)$ represents only a claimed prefix.
Storing a replay sequence in a certificate provides no guarantee of
validity. The certificate theorem applies only after accepted
evidence materializes a coherent, wellformed run that agrees with the
certificate's accessors.

\subsection{Evidence}\label{evidence}

An \textbf{evidence bundle} contains the source-side material the
checker inspects. This includes certificate binding evidence, the chunk plan,
chunk-read coordinates, chunk domains, observed row values or absences,
ordinary CDC observations, and the construction evidence for the raw
clean prefix.

The evidence bundle does not contain a destination-state proof. It does
not contain sink ACK semantics. It does not contain whole-table
claim-scope side conditions except where those are added later for the
whole-table specialization. The certificate layer is source-side.

\subsection{Materialized Run}\label{materialized-run}

The materialization relation $\Mat(C,E,R)$ connects certificate $C$ and
evidence $E$ to the model. Together, they describe the abstract DBLog
run $R$ consumed by the Layer~2 theorem. Materialization is
purely structural: it connects the evidence to an abstract run carrier
(specifying scope, frontier, clean prefix, chunk plan, chunk reads, and
CDC events) without asserting that the resulting run is wellformed
against the source history.

Concretely, $\Mat$ relates the pair to the evidence-determined run.
The materialized run's scope, frontier, clean prefix, chunk plan,
chunk-read observations, and CDC observations are fixed by what $E$
serializes and cannot be chosen arbitrarily by the checker. Evidence that does not
describe a run shape the checker supports is reported $\Unsupported$
instead of being accepted into the claim. The certificate's own
fields are tied to the materialized run by the coherence predicate
below. Materialization alone establishes neither wellformedness nor
observation fidelity. Those properties derive from the coherence predicate and
the checker obligations.

Rejected and unsupported evidence need not materialize a run. For
accepted evidence, the coherence predicate ensures that the checker
uses the run determined by the evidence.

\subsection{Certificate / Run
Coherence}\label{certificate-run-coherence}

We write $\Coh(C,E,R)$ for the certificate / run coherence predicate.
It packages materialization with accessor agreement between $C$ and
$R$:

\[
\begin{aligned}
\Coh(C,E,R) \,\Longleftrightarrow\,&
  \Mat(C,E,R)\\
  \land\,& \Scope(C)=\Scope(R)\\
  \land\,& \Frontier(C)=\Frontier(R)\\
  \land\,& \CleanPrefix(C)=\CleanPrefix(R).
\end{aligned}
\]

Coherence concerns the run materialized by this certificate and evidence
pair. It is stronger than the existence of some run with matching
accessors.

\subsection{Verifier Result}\label{verifier-result}

The verifier produces one of three outcomes
(Table~\ref{tab:verifier-results}):

\begin{table}[t]
\caption{Verifier result cases.}
\label{tab:verifier-results}
\centering\small
\begin{tabularx}{\columnwidth}{@{}lY@{}}
\toprule
\textbf{Result} & \textbf{Meaning}\\
\midrule
$\Accept$ &
  The certificate/evidence pair is internally coherent for the
  source-side checker fragment.\\
$\Reject$ &
  The evidence contradicts the certificate or fails a required check.\\
$\Unsupported$ &
  The evidence uses source-side vocabulary outside the checker fragment
  modeled here.\\
\bottomrule
\end{tabularx}
\end{table}

The $\Unsupported$ outcome prevents unsound acceptance when the
verifier encounters evidence outside its modeled fragment. If the checker
cannot interpret an optional source-side feature, the certificate is not
accepted as if that feature had been checked.

$\Accept$, $\Reject$, and $\Unsupported$ are verifier
results, not theorem-status labels. The verifier result here is an
abstract source-side checker result, not a claim that a serialized
certificate format, parser, or production verifier implementation has
been proved correct.

\subsection{What Accept Means, and What It Does Not
Mean}\label{what-accept-means-and-what-it-does-not-mean}

$\Accept$ indicates that, under accurate source observation, the
evidence materializes a wellformed DBLog run whose scope, frontier, and
clean prefix match the certificate. That provides the bridge back to
Layer 2.

$\Accept$ does not validate the concrete database, connector, log,
primary-key interpretation, source-coordinate semantics, parser, wire
format, or named product. It also does not show that a sink applied the
prefix or that the claim covers the whole table. Accurate source
observation remains an explicit assumption.

\subsection{Accurate Source
Observation}\label{faithful-source-observation}

We write $\FSO(E,b_0,H,\RawObs)$ for the accurate source observation
assumption ($\FSO$), the External observation assumption of the certificate
layer. It ties the finite evidence bundle $E$ to the mathematical
source history $H$ and base state $b_0$, against an abstract observation
carrier $\RawObs$. It covers the following source facts:

\begin{itemize}\tightlist
\item the source history and base state the certificate is about.
\item the meaning and ordering of source-log coordinates.
\item CDC retention and completeness over the certified interval.
\item preservation of same-coordinate order and multiplicity where
  relevant.
\item accuracy of chunk-read values and absences at their recorded
  coordinates.
\item the primary-key interpretation and the assumption that chunk
  reads observe committed source state.
\end{itemize}

The theorem statements write the raw source-observation carrier as
$\RawObs$. The carrier is abstract. It stands for a deployment's record
of source observations, interpreted under the accurate-source-observation
assumption ($\FSO$). It is not a
destination log, not a sink-state proof, and not a claim that a named
connector has been verified.

The certificate checker can compare evidence fields to one another. It
cannot prove, from the certificate alone, that the outside world
produced accurate observations. Consequently, the accepted-certificate
theorem takes accurate source observation as an explicit External observation
assumption.

\subsection{Role of the Certificate Layer}\label{why-this-layer-exists}

Layer~2 needs a wellformed mathematical run, while a deployment exposes
a certificate and evidence. Layer~3 connects these levels by establishing that
accepted evidence materializes a coherent wellformed run under accurate
source observation, enabling the application of the source-side virtual-cut
theorem.

\section{Accepted-Certificate Soundness \mbox{(Layer 3)}}\label{accepted-certificate-soundness-layer-3}

Layer 3 establishes the certificate-side theorem across two components: an
intermediate bridge theorem and the composed virtual-cut theorem.

\subsection{Intermediate Theorem}\label{intermediate-theorem}

The intermediate theorem is machine-checked under a small set of primitive
\emph{checker obligations} that the abstract certificate checker must meet
to guarantee a valid run. Table~\ref{tab:checker-substrate} lists them.
They are stated as explicit assumptions. We do not prove that a concrete
parser, wire format, or production verifier discharges them.

\begin{table}[t]
\caption{Checker obligations S1--S4 for the
accepted-certificate intermediate theorem.}
\label{tab:checker-substrate}
\centering\small
\begin{tabularx}{\columnwidth}{@{}l@{\hskip 1em}Y@{}}
\toprule
\textbf{} & \textbf{Checker obligation}\\
\midrule
S1 & Accepted inputs are wellformed checker inputs that are structurally
     valid and parseable.\\
S2 & Accepted checker inputs materialize an abstract DBLog run.\\
S3 & The materialized run's accessors (scope, frontier, and clean
     prefix) agree with the certificate.\\
S4 & Under accurate source observation, an accepted materialization is a
     wellformed run against the source history.\\
\bottomrule
\end{tabularx}
\end{table}

Obligations S1--S4 form the assumptions of the intermediate
theorem. Each is a property that the checker design must satisfy before
an $\Accept$ result implies the theorem's source-side conclusion.
Reporting unsupported evidence is not part of these formal assumptions,
but rather an operational checker obligation.

\smallskip
\noindent\textbf{Unsupported reporting.} A checker meeting S1--S4 must
not accept uninterpretable evidence as if it had been verified.
Such evidence is reported as $\Unsupported$ rather than
$\Accept$. \textbf{Status:} Deployment obligation on the concrete
checker, not a Layer~3 assumption.

Under obligations S1--S4, the intermediate theorem states:

\[
\begin{aligned}
&\Verify(C,E)=\Accept \,\land\, \FSO(E,b_0,H,\RawObs)\\
&\quad\Longrightarrow\, \exists R.\;
   \WF(b_0,R,H) \,\land\, \Coh(C,E,R).
\end{aligned}
\]

In words, accepted evidence, when interpreted under accurate source
observation, certifies a wellformed abstract DBLog run whose accessors
agree with the certificate. Obligations S1--S3 guarantee materialization
and accessor agreement, while S4 ensures wellformedness. The
unsupported-reporting requirement prevents uninterpretable evidence
from being treated as verified. This is an operational obligation rather
than a formal assumption.

\subsection{Composed Theorem}\label{composed-theorem}

The main certificate theorem is:

\[
\begin{aligned}
&\Verify(C,E)=\Accept \,\land\, \FSO(E,b_0,H,\RawObs)\\
&\quad\Longrightarrow\, \VC(C,H).
\end{aligned}
\]

Expanding this definition:

\[
\begin{aligned}
&\Verify(C,E)=\Accept \,\land\, \FSO(E,b_0,H,\RawObs)\\
&\quad\Longrightarrow
  \Apply(\CleanPrefix(C)) \restrict \Scope(C)\\
&\qquad= \Src(b_0,H,\Frontier(C)) \restrict \Scope(C).
\end{aligned}
\]

\paragraph*{Theorem Summary.}
\begin{itemize}
\tightlist
\item
  \textbf{Status.} Machine-checked under
  checker obligations S1--S4 (Table~\ref{tab:checker-substrate})
  and accurate source observation ($\FSO$).
\item
  \textbf{Claim.} An accepted certificate guarantees a virtual cut
  when its evidence is interpreted through accurate source observation.
\item
  \textbf{Assumptions.} Checker obligations S1 (input
  wellformedness), S2 (materialization), S3 (accessor agreement),
  S4 (accurate materialization is wellformed), and accurate source
  observation ($\FSO$).
\item
  \textbf{Boundary.} Acceptance does not prove a concrete parser, wire
  format, production verifier, named product, whole-table claim, or sink
  application. The unsupported-reporting requirement is a separate
  checker-design obligation rather than a formal assumption.
\end{itemize}

Consequently, certificate acceptance guarantees a virtual cut only relative
to accurate source observation. Acceptance alone cannot validate
physical database behavior or external connector fidelity.

\subsection{Proof Explanation}\label{proof-explanation}

The proof composes the layers already introduced.

First, verifier acceptance under accurate source observation establishes a
valid run $R$:

\[
\begin{aligned}
&\WF(b_0,R,H),\\
&\Coh(C,E,R).
\end{aligned}
\]

Second, Layer 2 applies to that wellformed run:

\[
\begin{aligned}
&\Apply(\CleanPrefix(R)) \restrict \Scope(R)\\
&\qquad= \Src(b_0,H,\Frontier(R)) \restrict \Scope(R).
\end{aligned}
\]

Third, coherence substitutes certificate accessors for run accessors:

\[
\begin{aligned}
\Scope(C)&=\Scope(R),\\
\Frontier(C)&=\Frontier(R),\\
\CleanPrefix(C)&=\CleanPrefix(R).
\end{aligned}
\]

After substitution, the Layer 2 equality becomes:

\[
\begin{aligned}
&\Apply(\CleanPrefix(C)) \restrict \Scope(C)\\
&\qquad= \Src(b_0,H,\Frontier(C)) \restrict \Scope(C).
\end{aligned}
\]

Fourth, by definition of $\VC(C,H)$ (Section~\ref{virtual-cuts}),
this yields the certified virtual cut.

The proof uses no sink, whole-table, or named-product assumption. Its
conclusion is source-side and restricted to $\Scope(C)$. The
checker obligations and accurate source observation remain
assumptions.

\subsection{Non-Vacuity and Boundary
Cases}\label{non-vacuity-and-boundary-cases}

The formal development includes the following non-vacuity checks and
boundary cases:

\begin{itemize}\tightlist
\item A non-trivial accepted certificate exists with a
  non-base frontier, a chunk read, a value-changing replay event, and
  valid $\RawObs$, $b_0$, and $H$ evidence.
\item candidate shapes lacking required proof components violate the
  checker obligations.
\item mismatched certificate-evidence pairs are rejected.
\item unknown source-side evidence is $\Unsupported$ rather than
  silently accepted.
\item mismatched certificate and run accessors fail coherence.
\item inaccurate $\RawObs$, mismatched $b_0$, or corrupted $H$ fail
  accurate source observation.
\item an empty scope satisfies the predicate vacuously, whereas the formal
  development provides a non-trivial concrete instance with non-empty keys.
\end{itemize}

These cases protect the main trust boundary. The theorem does not say
that the certificate eliminates trust. It states that, under accurate
source observation and explicit checker obligations, acceptance establishes a
coherent wellformed run. The source-side theorem then
yields the certified virtual cut.

\section{Whole-Table Specialization \mbox{(Layer 4)}}\label{whole-table-specialization-layer-4}

Layers~2 and~3 prove equality on $\Scope(C)$. In a full backfill,
this scope is typically intended to cover an entire table or partition.
Such a claim needs additional evidence that the scope includes every
key whose state can matter between an anchor boundary and the certified
frontier. The replay must also avoid writing keys outside that scope.

Whole-table equality is therefore a derived result rather than an
inherent guarantee of certificate acceptance. Acceptance yields a
scoped virtual cut, and the whole-table side condition establishes when
the restriction to $\Scope(C)$ can safely be removed.

\subsection{Anchor Boundary and Anchor
Domain}\label{anchor-boundary-and-anchor-domain}

An anchor boundary $b$ is a source coordinate used to define the key
population that the whole-table claim covers. It is not a
physical snapshot read. Chunks may be read after $b$. Stale chunk observations can
be repaired by source-log evidence and replayed to the frontier.

For a certificate $C$, the anchor must not be later than the
certified frontier:

\[
b \le \Frontier(C).
\]

The anchor domain is the set of keys present in the source at that
boundary:

\[
  \AnchorDomain(H,b)
    = \{\,k \mid \Src(b_0,H,b)(k) \text{ is present}\,\}.
\]

The anchor domain inherits its dependence on the base state $b_0$ from
$\Src$. As with $\Src$, $b_0$ is a single base state fixed throughout the
paper. We keep it implicit in $\AnchorDomain(H,b)$, and likewise in the
whole-table claim scope $\WTS(C,H,b)$ below, since it never varies. A
key present at the anchor that no source event at or before $b$ ever
writes inherits its presence entirely from $b_0$, the only place where the
base state directly affects the whole-table claim.

Operationally, a whole-table deployment may use a maximum-primary-key
query or equivalent end-of-table indicator to show that the claimed scope
covers this anchor domain. Such an indicator does not replace $\WTS$. It
must be tied to the anchor boundary $b$, or otherwise justify
$\AnchorDomain(H,b) \subseteq \Scope(C)$. It also does not discharge the
post-anchor touch condition below, including for keys beyond an
anchor-time maximum.

Keys absent at $b$ can still become relevant later. If the source
inserts, updates, or deletes a key after $b$ and at or before the
certified frontier, the whole-table claim must account for that key too.

$\TouchedBetween(H,b,f)$ is the set of keys $k$ for which $H$
has an insert, update, or delete at some source coordinate $c$
with $b < c \le f$, and the conservative interval table scope is
\[
\begin{aligned}
&\TableScopeBetween(H,b,f)\\
&\quad= \AnchorDomain(H,b) \cup \TouchedBetween(H,b,f).
\end{aligned}
\]

This event-touch scope is conservative rather than a minimal
semantic-difference set: even deleting an already absent key counts as
a touch that the claim scope must include.

\subsection{A Small Anchor Example}\label{a-small-anchor-example}

Suppose the source contains keys $100$ and $200$ at anchor
boundary $b$. Between $b$ and frontier $f$, key
$300$ is inserted and key $200$ is deleted.

\[
\begin{aligned}
\AnchorDomain(H,b)&=\{100,200\},\\
\TouchedBetween(H,b,f)&=\{200,300\},\\
\TableScopeBetween(H,b,f)&=\{100,200,300\}.
\end{aligned}
\]

A whole-table claim for this interval must include all three keys in
$\Scope(C)$. Key $100$ is present at the anchor. Key
$200$ is present at the anchor and then deleted. Key $300$
is absent at the anchor but inserted before the frontier. Omitting any
of them would make the whole-table statement stronger than the evidence
supports.

There is also a replay-side condition. If the certified clean prefix
writes a key outside $\Scope(C)$, such as key $999$, scoped equality
may still hold, but the no-unclaimed-replay-key condition fails. The
whole-table theorem therefore does not establish unrestricted equality
for that prefix.

\subsection{Whole-Table Claim Scope}\label{whole-table-claim-scope}

We write $\WTS(C,H,b)$ for the whole-table claim-scope side condition,
defined by four conjuncts:

\[
\begin{aligned}
&\WTS(C,H,b) \,\Longleftrightarrow\, b \le \Frontier(C)\\
&\,\land\, \AnchorDomain(H,b) \subseteq \Scope(C)\\
&\,\land\, \TouchedBetween(H,b,\Frontier(C)) \subseteq \Scope(C)\\
&\,\land\, \ReplayKeys(\CleanPrefix(C)) \subseteq \Scope(C).
\end{aligned}
\]

The first conjunct fixes the ordering of the anchor and frontier. The
next two require $\Scope(C)$ to cover every key present at the
anchor and every key touched between the anchor and the frontier. The
final conjunct ensures that the certified replay prefix writes no
unclaimed keys.
Figure~\ref{fig:whole-table-claim-scope} illustrates the four conjuncts as
overlapping key regions (anchor-domain, touched-between, replay-keys)
over the source-coordinate axis.

\begin{figure*}[t]
  \input{figures/fig05_anchor_domain_table_scope/fig05.tex}
\end{figure*}

This predicate is explicit because whole-table correctness is not
automatic. Certificate acceptance alone gives a scoped statement. The
whole-table side condition is the extra bridge from ``correct on the
certified claim scope'' to ``correct on every table key''.

\subsection{Whole-Table Theorem}\label{whole-table-theorem}

Under certificate acceptance, accurate source observation, and the
whole-table claim-scope side condition, the certified clean prefix
reaches the source state at the certified frontier on the whole table:

\[
\begin{aligned}
&\Verify(C,E)=\Accept \,\land\, \FSO(E,b_0,H,\RawObs)\\
&\,\land\, \WTS(C,H,b)\\
&\,\Longrightarrow\, \Apply(\CleanPrefix(C))=\Src(b_0,H,\Frontier(C)).
\end{aligned}
\]

\paragraph*{Theorem Summary.}
\begin{itemize}
\tightlist
\item
  \textbf{Status.} Machine-checked under
  checker obligations S1--S4
  (Table~\ref{tab:checker-substrate}), accurate source observation ($\FSO$),
  and $\WTS(C,H,b)$.
\item
  \textbf{Claim.} The scoped certificate theorem lifts to whole-table
  source-state equality when the whole-table claim-scope side conditions
  cover boundary order, anchor-domain keys, post-anchor touches, and
  replay keys.
\item
  \textbf{Assumptions.} Certificate acceptance, accurate source
  observation, $\WTS(C,H,b)$, and checker obligations
  S1--S4 (Table~\ref{tab:checker-substrate}).
\item
  \textbf{Boundary.} This remains a source-side theorem, distinct from sink
  convergence, exactly-once delivery, or arbitrary destination
  correctness.
\end{itemize}

This is a source-side theorem. The certificate checker and
accurate-source-observation assumption supply the certified scoped
virtual cut. The whole-table side condition supplies the coverage
needed to remove the scope restriction.

\subsection{Proof Outline}\label{proof-roadmap}

Layer 3 gives the scoped equality:

\[
\begin{aligned}
&\Apply(\CleanPrefix(C)) \restrict \Scope(C)\\
&\,= \Src(b_0,H,\Frontier(C)) \restrict \Scope(C).
\end{aligned}
\]

For keys inside $\Scope(C)$, that is already the desired equality.
For keys outside $\Scope(C)$, the whole-table side condition addresses both
source and replay states via the inclusions:
\[
  \begin{aligned}
    \AnchorDomain(H,b)                &\subseteq \Scope(C), \\
    \TouchedBetween(H,b,\Frontier(C)) &\subseteq \Scope(C), \\
    \ReplayKeys(\CleanPrefix(C))      &\subseteq \Scope(C)
  \end{aligned}
\]
By contraposition, any key $k \notin \Scope(C)$ belongs neither to $\AnchorDomain(H,b)$,
nor to $\TouchedBetween(H,b,\Frontier(C))$, nor to $\ReplayKeys(\CleanPrefix(C))$.
First, because $k \notin \AnchorDomain(H,b)$, it is absent at the anchor
boundary (even with base state $b_0$ taken into account). Because
$b \le \Frontier(C)$ and $k$ is not touched in $(b, \Frontier(C)]$,
$k$ remains absent from the source at the frontier.
Second, because $k \notin \ReplayKeys(\CleanPrefix(C))$, the clean prefix
never writes $k$. Replay and source are therefore both absent outside
$\Scope(C)$. Combining the inside-scope and outside-scope cases gives
unrestricted equality.

\subsection{Scope of the Result}\label{boundaries}

The specialization does not change the paper's trust boundaries,
continuing to rely on accurate observation of source-side evidence. It
does not prove that a concrete parser or production verifier
is correct, nor does it establish sink-state convergence.
It also distinguishes a virtual cut from a physical cut. The anchor provides a
source-coordinate boundary for reasoning about table scope rather than representing a
single request-wide read timestamp.

\section{Source-Side Continuation and Restriction of Virtual
Cuts}\label{source-side-continuation-and-restriction}

The results so far fix replay equality at one frontier. After the chunk
loop closes, DBLog continues consuming source-log events. We now show
how the established equality advances with that stream.

Continuation across frontiers appends the source changes committed
after frontier $f$ and through a later frontier $f'$. Restriction shows
that a virtual cut on scope $K$ remains valid on every smaller scope.
Both results are source-side and machine-checked. They contain no
destination, replica, or delivery term. Disjoint composition,
verification of extended certificates, and normalization of overlapping
or retried chunks are not proved here.

Both proofs proceed per key. Continuation uses the Layer~0 fact that the
source state at $f'$ is determined by the state at $f$ together with the
events in $(f,f']$. Restriction follows directly from equality on each
key in the original scope.

\subsection{Continuation Segments}\label{continuation-segments}

For a source history $H$, a key scope $K$, and source frontiers
$f \le f'$, the continuation segment for the half-open interval
$(f,f']$ on $K$ is obtained by filtering $H$. We keep every source event
whose coordinate $c$ satisfies $f<c\le f'$ and whose key lies in $K$.
The events remain in source order and are read as CDC replay events. We
write $\ContSeg(H,K,f,f',\delta)$ when $\delta$ is exactly this list,
with source order and multiplicity preserved.

By construction, $\ContSeg(H,K,f,f',\delta)$ requires $f \le f'$. The
predicate is false whenever $f > f'$, independent of $\delta$. Theorems
whose assumptions include $\ContSeg(H,K,f,f',\delta)$ therefore implicitly
require $f \le f'$. The order is not restated in their displayed
formulas.

The interval is half-open because coordinate indexing is inclusive
(Section~\ref{source-histories-and-frontiers}). Because $\Src(b_0,H,f)$ already
incorporates all events through coordinate $f$, continuation begins strictly
after $f$ and extends through $f'$ inclusive.

Because the segment is defined by filtering the source history, a list
that reorders, drops, or duplicates a source event does not satisfy
$\ContSeg(H,K,f,f',\delta)$. A production CDC feed must deliver this
exact sequence. Its accuracy is an \textbf{External observation
assumption}, and retaining the source-log interval $(f,f']$ is a
\textbf{Deployment obligation}.

The segment is filtered to the scope $K$. It is the change data for
the keys of $K$, not the source's entire change stream. A deployment
whose CDC feed includes all keys must project that stream onto $K$ to
obtain the continuation segment for a $K$-scoped cut. For a
whole-table continuation the scope is the universal key set, and the
whole-table continuation result below is the form to use.
This definition is strict: omitting, reordering, or duplicating an
\emph{in-scope} event violates $\ContSeg(H,K,f,f',\delta)$, invalidating
the continuation. Padding the segment with \emph{out-of-scope} events
also fails $\ContSeg$, even though the resulting equality constrains
only keys in $K$.

\subsection{Continuation Across
Frontiers}\label{continuation-across-frontiers}

The primary theorem of this section continues a virtual cut to a later
frontier. Let $H$ be a wellformed source history. If $\sigma$ is a
virtual cut at frontier $f$ on scope $K$, and $\delta$ is the
continuation segment for $(f,f']$ on $K$, then $\sigma$ followed by
$\delta$ (sequence concatenation, written $\sigma \mathbin{@} \delta$)
is a virtual cut at the later frontier $f'$ on the same scope:

\[
\begin{aligned}
&H \text{ wellformed} \,\land\, \VCS(\sigma,K,f,H)\\
&\,\land\, \ContSeg(H,K,f,f',\delta)\\
&\,\Longrightarrow\, \VCS(\sigma \mathbin{@} \delta, K, f', H).
\end{aligned}
\]

\paragraph*{Theorem Summary.}
\begin{itemize}
\tightlist
\item
  \textbf{Status.} Machine-checked under the displayed
  assumptions.
\item
  \textbf{Claim.} Appending the valid continuation segment for
  $(f,f']$ advances a virtual cut at $f$ to a virtual cut at $f'$ on the
  same scope.
\item
  \textbf{Assumptions.} $H$ is a wellformed source history, $\sigma$ is a
  virtual cut at $f$ on $K$, and $\delta$ is exactly the continuation
  segment for $(f,f']$ on $K$.
\item
  \textbf{Boundary.} The conclusion is a replay-against-source equality
  on $K$. It is not a whole-table claim, not a statement that a verifier
  accepts an extended certificate, and not a destination-apply claim.
\end{itemize}

The wellformedness condition is needed because the proof uses the
source history's coordinate order to know that the events in $(f,f']$
are exactly the ones between the two frontiers. Without monotonicity,
events outside the interval could interleave.

\emph{Proof idea.} The argument is pointwise on a key $k \in K$. Replay
of a sequence at $k$ is fixed by the latest replay event for $k$. If the
continuation segment $\delta$ contains an event for $k$, then the latest
$k$-event of $\sigma \mathbin{@} \delta$ lies in $\delta$ and is the
change-data reading of the latest source event for $k$ in $(f,f']$. A
Layer~0 source-state interval lemma equates that event's effect with
$\Src(b_0,H,f')$ at $k$. If $\delta$ contains no event for $k$, replay of
$\sigma \mathbin{@} \delta$ at $k$ equals replay of $\sigma$ at $k$,
which the hypothesis equates with $\Src(b_0,H,f)$ at $k$. The same interval
lemma equates that with $\Src(b_0,H,f')$ at $k$, because no event for
$k$ in the interval means the source state for $k$ does not change
across it. The interval lemma (that $\Src$ at $f'$ is determined by
$\Src$ at $f$ together with the source events in $(f,f']$) is the one
new Layer~0 fact the continuation needs. It is proved from the
source-coordinate order.

\smallskip
\noindent\textbf{Whole-table continuation.} The whole-table
specialization (Section~\ref{whole-table-specialization-layer-4})
delivers, under its claim-scope side conditions, the \emph{unrestricted}
equality $\Apply(\sigma) = \Src(b_0,H,f)$. Continuation specializes to
that case. For a wellformed $H$, if $\Apply(\sigma) = \Src(b_0,H,f)$
holds on every key and $\delta$ is the continuation segment for
$(f,f']$ on every key, then
$\Apply(\sigma \mathbin{@} \delta) = \Src(b_0,H,f')$. This is the
all-keys instance of the theorem above, machine-checked under the same
assumptions. A \emph{scoped} continuation cannot be promoted to a
whole-table guarantee without this all-keys condition. A key the source
inserts outside $K$ within $(f,f']$ changes the table while leaving the
scoped equality on $K$ intact. Whole-table continuation requires the
continuation segment for every key, just as whole-table specialization
requires its claim-scope side condition.

\subsection{Restriction to a
Sub-Scope}\label{restriction-to-a-sub-scope}

A virtual cut also restricts downward along the scope. If $\sigma$ is a
virtual cut at frontier $f$ on scope $K$, then it is a virtual cut at
$f$ on every sub-scope $K' \subseteq K$:

\[
\VCS(\sigma,K,f,H) \,\land\, K' \subseteq K
  \,\Longrightarrow\, \VCS(\sigma,K',f,H).
\]

This is machine-checked. It holds for any sub-scope $K' \subseteq K$.
Restriction is strictly one-directional. \emph{Widening} a scope is
unsound. For a key outside $K$ the cut certifies nothing about that
key's state at $f$, so there is no sound lemma extending a cut from $K$
to a larger scope.

The same fact holds through a certificate's accessors. If $C$ is a
certificate with $\VC(C,H)$ and $K' \subseteq \Scope(C)$, then the
certified clean prefix is a virtual cut at $\Frontier(C)$ on the
narrowed scope $K'$:

\[
\begin{aligned}
&\VC(C,H) \,\land\, K' \subseteq \Scope(C)\\
&\quad\Longrightarrow\,
  \VCS\bigl(\CleanPrefix(C),K',\Frontier(C),H\bigr).
\end{aligned}
\]

This restricts the equality already established by a certificate to a
smaller scope. Validating a restricted certificate for $K'$ requires
constructing and checking the corresponding evidence bundle, rather than
following automatically from the restriction lemma alone.

\subsection{Continuation Over an Accepted
Certificate}\label{continuation-over-an-accepted-certificate}

Continuation composes with the certificate layer. Layer~3
(Section~\ref{accepted-certificate-soundness-layer-3}) proves that an
accepted certificate, under accurate source observation, certifies a
virtual cut at $\Frontier(C)$ on $\Scope(C)$. Appending a continuation
segment advances that cut forward. Let $H$ be a wellformed source
history. If a certificate $C$ is accepted under accurate source
observation, and $\delta$ is the continuation segment for
$(\Frontier(C),f']$ on $\Scope(C)$, then the certified clean prefix
followed by $\delta$ is a virtual cut at $f'$ on $\Scope(C)$:

\[
\begin{aligned}
&\Verify(C,E)=\Accept \,\land\, \FSO(E,b_0,H,\RawObs)\\
&\,\land\, H \text{ wellformed}\\
&\,\land\, \ContSeg\bigl(H,\Scope(C),\Frontier(C),f',\delta\bigr)\\
&\,\Longrightarrow\, \VCS\bigl(\CleanPrefix(C) \mathbin{@} \delta, \Scope(C), f', H\bigr).
\end{aligned}
\]

\paragraph*{Theorem Summary.}
\begin{itemize}
\tightlist
\item
  \textbf{Status.} Machine-checked under
  checker obligations S1--S4
  (Table~\ref{tab:checker-substrate}), accurate source observation,
  source-history wellformedness, and exact continuation on
  $(\Frontier(C),f']$.
\item
  \textbf{Claim.} An accepted certificate's virtual cut continues, by
  appending a valid continuation segment, to a virtual cut at a later
  frontier.
\item
  \textbf{Assumptions.} Certificate acceptance under checker
  obligations S1--S4 (Table~\ref{tab:checker-substrate}) and accurate
  source observation, a wellformed source history, and a
  continuation segment matching the source's change data on
  $(\Frontier(C),f']$.
\item
  \textbf{Boundary.} The conclusion is a replay equality on the
  sequence $\CleanPrefix(C) \mathbin{@} \delta$. It is not a
  claim that the verifier accepts an extended certificate.
  Acceptance for the longer prefix remains an evidence-and-verifier
  obligation.
\end{itemize}

The continuation segment $\delta$ is itself an external observation of
the source. Its accuracy with respect to $H$ is an assumption that
must be satisfied at deployment, exactly as the certificate's own
evidence is. The theorem composes the Layer~3 certificate theorem with
continuation across frontiers. The composition leaves the trust
assumptions for the certificate evidence and the continuation segment
unchanged.

\subsection{Ongoing Operation as a Sequence of
Cuts}\label{ongoing-operation-as-a-sequence-of-cuts}

Continuation segments chain. For a wellformed source history, the
continuation segment for $(f,f']$ followed by the continuation segment
for $(f',f'']$ is the continuation segment for $(f,f'']$. Wellformedness
is again required. Only a non-decreasing source history guarantees that
the two coordinate ranges meet at $f'$ without interleaving. Given a
sequence of frontiers, the complete in-scope CDC segment over the combined
interval can be split into segments at those frontiers. Appending the
segments one at a time yields the same replay state as appending the
combined segment.

DBLog's replay-equivalence claim can therefore advance across frontiers.
Each later cut consists of the original cut followed by the source
changes committed in between.

\smallskip
\noindent\textbf{Re-chunking and re-running.} This property also governs
repeated runs. Because a virtual cut at $(K,f)$ depends solely on
$\Src(b_0,H,f)$, any two replay sequences forming virtual cuts at $(K,f)$
necessarily agree on $K$. A fresh wellformed run for the same scope and
frontier reconstructs the same source state on that scope, independently
of downstream replica state.
Re-chunking is therefore an absolute, source-determined reconstruction of
the per-key state (the regenerated replay prefix need not be identical,
but replaying it yields the same source state), rather than a differential
repair against a destination.
Whether a destination then applies the regenerated prefix is a
separate, sink-side question, which
Section~\ref{sink-state-convergence-non-claim} leaves outside the
current claims.

\subsection{Scope of the Results}\label{continuation-boundaries}

Each conclusion is a source-side $\Apply$-against-$\Src$ equality. The
continuation segment must be the source's real change data
on the interval, which is an External observation assumption with a
Deployment obligation to retain that interval. These results prove no
whole-table claim beyond the explicit all-keys instance, no acceptance of an
extended or restricted certificate, and no destination-state
convergence. Continuation and restriction are machine-checked. Broader
certificate-algebra extensions remain future work.

\section{Limits on Sink-State Convergence}\label{sink-state-convergence-non-claim}

Layers~2 through~4 are source-side. They relate the certified replay
prefix to the source at the certified frontier. They do not establish
the state of a downstream destination.

Three destination properties must be distinguished. An \emph{ACK}
records receipt. A \emph{durable audit log} persists what was received.
\emph{Applied state} is the state reached by the destination's tables.

\subsection{Receipt ACK}\label{receipt-ack}

A destination ACK confirms that data was received. It does not
guarantee that the destination applied that data, applied it in order,
executed it idempotently, processed it without duplicates, or converged
to the source state at the certified frontier. Receipt could support a
future stream-correctness result, but it does not establish sink-state
correctness.

\subsection{Durable Audit-Log
Persistence}\label{durable-audit-log-persistence}

A destination may durably record the certified clean prefix together
with a run id, scope, frontier, or digest. This gives an auditable record
of what arrived. Durable receipt is still
stream-level evidence, not state-level evidence. A log can contain the
right events while the destination table has not applied them or has
applied them under different semantics.

\subsection{Applied-State Evidence}\label{applied-state-evidence}

To prove destination state convergence, a future layer would need a sink
model with ordering, idempotence, fencing, visible apply barriers, and a way
to observe that the destination state after the barrier corresponds to
applying the certified clean prefix. This extension remains future work.

Even that future theorem would be scoped. Using $\SinkState$ as a
placeholder for a future sink-state accessor, the target statement has
the following form:

\[
\SinkState \restrict \Scope(C)
  = \Src(b_0,H,\Frontier(C)) \restrict \Scope(C).
\]

Whole-table sink convergence would additionally need the whole-table
side conditions from Section~10. Certificate acceptance alone does not
guarantee that the destination database is correct. Establishing sink-state
convergence requires both the source-side replay guarantee and an explicit
sink execution contract.

A virtual cut relates replay to the source. An ACK establishes
receipt, while an audit log establishes persistence. Destination-state
guarantees require a fenced, idempotent apply contract and visible barrier
evidence.

\section{Related Work}\label{related-work}

The closest work comes from change-data-capture platforms, distributed
cuts and frontiers, stream-processing watermarks, database recovery,
and formal certificates. We compare these areas by concept and use
named systems as deployment context.

\subsection{DBLog Mechanism Lineage and Production
Adopters}\label{dblog-mechanism-lineage-and-production-adopters}

The DBLog mechanism was introduced in the December 2019 Netflix Tech
Blog post~\citep{dblog_netflix_blog} and described operationally in the
2020 DBLog paper~\citep{andreakis2020dblog}. It scans a table in
primary-key chunks bracketed by log watermarks, discarding chunk rows
whose keys overlap with CDC events anywhere within the window and
emitting the remaining rows after the high watermark. We
formalize this reconstructed state as a certified virtual cut. The model uses a canonical,
non-overlapping chunk partition. The original mechanism's retried or
overlapping chunk reads must first be normalized into that canonical
form. That normalizer is not developed here
(Section~\ref{certificate-algebra-and-model-extensions}).

Debezium \citep{debezium_incremental_snapshot} and Apache Flink
CDC \citep{flink_cdc_incremental_snapshot} include DBLog-style
mechanisms in open-source CDC deployments. Debezium provides
incremental snapshot modes, including signaling-table watermarks and
read-only variants that use executed GTID sets for MySQL and MariaDB
and transaction-snapshot information for PostgreSQL
\citep{debezium_read_only_snapshots}.
Apache Flink CDC combines Debezium-style connectors with Flink's
checkpointed streaming runtime \citep{carbone2015asynchronous},
including parallel chunk readers \citep{flink_cdc_incremental_snapshot}.

The formal result does not transfer to these systems automatically. A
deployment must still establish source-log retention, watermark or
coordinate semantics, frontier ordering, and accurate source
observation.

\subsection{Cuts, Frontiers, Watermarks, and
Checkpoints}\label{cuts-frontiers-watermarks-and-checkpoints}

The term ``virtual cut'' follows the distributed-systems language of
logical time and consistent cuts. Lamport's logical clocks
\citep{lamport1978time}, Chandy-Lamport snapshots
\citep{chandy1985snapshots}, Mattern's virtual time
\citep{mattern1989virtual}, Fidge's vector clocks
\citep{fidge1988vector}, and Pu's on-the-fly consistent reading
\citep{pu1986consistentreading} are the nearest theoretical neighbors.

A DBLog virtual cut is not a global consistent cut over all processes.
It is an extensional replay-equality statement. Replaying a certified
prefix yields the source state at a frontier on a key scope.
Ordering is governed by source-history coordinates instead of a
vector-clock model of inter-process channels. Each cut is point-in-time,
fixed to one frontier, but the cuts chain. The source-side continuation
result (Section~\ref{source-side-continuation-and-restriction}) advances
a cut from one frontier to a later one, so a DBLog run corresponds to a
sequence of source-side virtual cuts rather than a single one.

Stream-processing systems use adjacent terms. The Dataflow model
\citep{akidau2015dataflow} popularized event time and watermarks. Naiad
\citep{murray2013naiad} uses frontiers as progress markers. Flink's
asynchronous snapshots \citep{carbone2015asynchronous} checkpoint
operator state for recovery. DBLog's watermarks are source evidence
around chunk reads, and $\Frontier(C)$ is the source-history
boundary for replay equality. It is not a stream-processing checkpoint
of operator state, and it is not an exactly-once processing claim.

\subsection{CDC Infrastructure and Source-Log
Semantics}\label{cdc-infrastructure-and-source-log-semantics}

Data ingestion, transport, and coordination systems such as LinkedIn Databus
\citep{databus}, Facebook Wormhole \citep{wormhole}, Kafka
\citep{kafka}, Kafka Connect JDBC \citep{kafka_connect_jdbc}, and
ZooKeeper \citep{zookeeper} provide infrastructure for copying,
transporting, and coordinating data. Kafka Connect JDBC provides
polling-based table ingestion. These systems do not provide the
certificates described in this paper.

Source-database documentation for MySQL binary logs and GTIDs
\citep{mysql_binlog}, PostgreSQL logical decoding
\citep{postgres_logical_decoding}, PostgreSQL logical replication
\citep{postgres_logical_replication}, and MariaDB GTIDs
\citep{mariadb_binlog} identifies the source-log ordering and row-event
semantics that a deployment must establish for accurate source
observation.

Database-native and cloud CDC systems include PostgreSQL logical replication
\citep{postgres_logical_replication}, CockroachDB changefeeds
\citep{cockroachdb_changefeeds}, AWS DMS \citep{aws_dms}, GCP Datastream
\citep{gcp_datastream}, Oracle GoldenGate \citep{oracle_goldengate},
Striim \citep{striim}, Materialize's PostgreSQL ingestion
\citep{materialize_postgres}, Databricks Lakeflow
\citep{databricks_lakeflow}, and Microsoft SQL Server CDC
\citep{microsoft_sqlserver_cdc}. These systems provide components or
workflows for the broader ``initial load plus change stream'' family. Our result adds a finite
replay-equivalence certificate for the DBLog form of this pattern,
under explicit source-observation assumptions.

CockroachDB resolved timestamps provide a production analogue of a
source frontier. PostgreSQL logical replication illustrates the
initial-copy-plus-ordered-change-stream pattern. Neither is treated here
as exporting the finite, scoped evidence required by a virtual-cut
certificate.

\subsection{Recovery, Snapshots, Isolation, and Live
Backfill}\label{recovery-snapshots-isolation-and-live-backfill}

A database snapshot may refer to a recovery snapshot, a multi-version
concurrency control (MVCC) read view, or point-in-time recovery (PITR).
ARIES \citep{mohan1992aries} and PostgreSQL PITR
\citep{postgres_pitr} reconstruct database state from checkpoints and
logs. CALC \citep{ren2016calc} shows that database systems can construct
logically consistent checkpoints without conventional synchronous
physical checkpointing. Our contribution is different. It provides a
scoped, source-side replay-equivalence certificate for a DBLog backfill,
whose clean replay prefix equals source state at a named frontier and
can be advanced to later frontiers. Berenson et al.
\citep{berenson1995isolation}, Adya et al. \citep{adya2000isolation},
and Oracle Flashback \citep{oracle_flashback} belong to the
transaction-isolation and as-of-read tradition.

DBLog is not a recovery algorithm, an online backup mechanism, an MVCC
snapshot, or a snapshot-isolation transaction. It does not assert that
all chunk reads happened at one source timestamp. It asserts replay
equality at a chosen frontier on a chosen scope.

Online schema migration tools such as gh-ost \citep{ghost2016}, F1
online schema change \citep{rae2013f1}, and lazy schema migration
systems \citep{zeng2024slsm} share the operational pressure of copying or
rewriting data while writes continue. Their overlap with DBLog is live
backfill under concurrent writes. Our formal result specifically
establishes source-side replay equality for a DBLog-style backfill and
its frontier advancement, rather than schema-state transitions or table
swaps.

\subsection{Related Verification Approaches}\label{sec:related-verification-approaches}

The certificate layer follows the general idea of checkable evidence
under explicit assumptions. Refinement mappings \citep{abadi1991refinement},
proof-carrying code \citep{necula1997pcc}, translation validation
\citep{pnueli1998translation}, version certificate recovery
\citep{clark2024vcr}, and VerIso \citep{ghasemirad2025veriso} all help
position the idea of a finite object that establishes correctness relative
to a model. Large mechanized artifacts such as seL4
\citep{klein2009sel4}, CompCert \citep{leroy2009compcert}, TLA+
\citep{lamport2002specifying}, Verdi \citep{wilcox2015verdi}, IronFleet
\citep{hawblitzel2015ironfleet}, and Alloy \citep{jackson2002alloy} are
cited for methodological comparison, not as theorem sources.

A DBLog certificate is not self-authenticating proof-carrying code, not
a sink refinement map, and not an isolation-level conformance
certificate. It is checkable evidence that, under accurate source
observation, establishes a wellformed DBLog run and therefore certifies a virtual
cut.

\subsection{Scope of the Comparison}\label{title-promise-and-product-claim-boundaries}

``Snapshot-equivalent replay'' is intended literally but narrowly. A
DBLog virtual cut is not a physical snapshot read, not a database
snapshot-isolation read view, not PITR, not a stream-processing
checkpoint, not a product replication guarantee, and not sink-state
correctness. The result is a snapshot-equivalent replay certificate that
does not require one physical snapshot read.

Debezium, Apache Flink CDC, MySQL, PostgreSQL, MariaDB, and the other
named systems provide mechanism or deployment context. Applying the
theorem to a concrete deployment requires its source-side assumptions,
evidence semantics, and checker obligations to be established.

\section{Future Work}\label{future-work}

Section~\ref{source-side-continuation-and-restriction} proves
continuation across frontiers and restriction to a sub-scope. The
following extensions are not proved here.

\subsection{Concrete Deployment
Paths}\label{sec:concrete-deployment-paths}

Extending the model to Debezium's read-only modes would replace source
watermark writes with observed markers: executed GTID sets for MySQL and
MariaDB, and transaction-snapshot information for PostgreSQL
\citep{debezium_read_only_snapshots}. A future theorem would need a
refinement showing that these markers establish the same clean-prefix
obligations as written watermarks, under the corresponding source
observation assumptions.

Apache Flink CDC parallel chunk execution would require a parallel
chunk-plan model, per-worker recovery assumptions, and a theorem
connecting checkpointed parallel execution to the same abstract
wellformed run used here. The certificate would need enough evidence to
recover the single clean prefix consumed by the proof.

SELECT-at-LSN and source-positioned read modes would replace low/high
watermark intervals with a point coordinate at which
a chunk read's row image is known to hold. These modes are attractive
because they can avoid source writes, but they need new source-side
assumptions and a new verification model for point-coordinate validation.

\subsection{Sink Extensions}\label{sink-extensions}

Destination-side results require a separate sink model.

ACK-only stream correctness would study receipt of the certified
stream prefix. An ACK can support a theorem about receipt, but not about
applied destination state. Such a proof would need a sink-receipt model
and a stream-prefix equality theorem.

Strong sink barrier convergence would establish applied destination
state. It needs a separate fenced and idempotent sink
contract, visible apply-barrier evidence, and a proof that the
destination state after the barrier realizes the certified clean
prefix. It is not a consequence of the accepted-certificate theorem
or the whole-table specialization by itself.

\subsection{Certificate Algebra and Model
Extensions}\label{certificate-algebra-and-model-extensions}

We prove restriction to a sub-scope and continuation across later
frontiers by valid CDC segments. Disjoint composition, concrete
extended-certificate verification, and a raw overlap/retry normalizer
remain open problems. The current Layer 1 model uses a strict canonical
chunk partition. A future normalizer would take overlapping or retried
raw chunk reads and produce a canonical strict-partition certificate plus
residual obligations, then prove that acceptance after normalization
preserves the same virtual-cut claim.

Some notation and layer-boundary choices also remain open. A
future paper might expose $\CleanPrefix(C,E)$ instead of
$\CleanPrefix(C)$, use a relation form such as
$\mathit{certifiesPrefix}$, or move overlap/retry concepts into Layer 1
instead of treating them through certificate materialization. Those
changes would alter the public theorem interface and require an explicit
revision.

\section{Verification Methodology}\label{verification-methodology}

The companion Isabelle/HOL development checks the source and replay model,
wellformed-run soundness, accepted-certificate soundness, whole-table
specialization, continuation, and restriction. Each theorem holds under
the assumptions stated with it.

A deployment must establish the assumed source-log retention, coordinate
semantics, chunk coverage, and accuracy of CDC and chunk observations.
The paper explains the definitions and proof arguments. Isabelle/HOL
checks the formal derivations under their assumptions.

Constructed models establish non-vacuity, including a non-trivial
accepted-certificate instance. Counterexample fixtures illustrate failures
when selected conditions are removed. Some conjuncts follow from the
construction or other conditions, as documented in the formal development.

\section{Conclusion}\label{conclusion}

In this paper, we formalized how DBLog combines chunk reads with captured
changes and proved the correctness of the resulting replay.
DBLog copies a table or a selected key range by reading
data in chunks while the source continues accepting writes. Because chunks are 
read at different source-log positions, the collected rows do not form a physical 
snapshot.

We formalized this reconstructed state through the concept of a \textbf{virtual cut}. 
A virtual cut establishes snapshot equivalence. Replaying the sequence of chunk 
observations and streaming change events produces the exact state of the source database at 
a designated commit-log frontier, for every key in scope. The cut is \emph{virtual} because 
rows are read at different times without pausing updates or taking a physical snapshot.
DBLog emits the events needed to reconstruct the source state downstream,
without storing that state itself.

We prove that this equivalence holds for every protocol run satisfying 
the stated run conditions. We then lift this guarantee 
to inspectable certificates, showing that an independent checker can verify 
the cut directly from recorded log evidence. When chunk reads cover the full 
table, this per-key guarantee establishes whole-table snapshot equivalence, 
accounting for both baseline rows and concurrent writes. We also prove that 
appending subsequent CDC events advances an established cut forward to later 
frontiers, and that cuts remain valid when restricted to smaller scopes. All 
results are machine-checked in Isabelle/HOL.

A certificate can expose missing or inconsistent evidence, but it
cannot prove that a connector's observations accurately reflect the
physical source database. Source-log retention, coordinate semantics,
frontier selection, and chunk-read accuracy remain deployment
obligations. The theorems establish source-side replay equivalence. 
They do not assert downstream sink convergence or exactly-once delivery, 
leaving concrete sink engines to separate verification.

\section*{Formal Development and Tools}\label{formal-development-and-tools}

The companion Isabelle/HOL development machine-checks the definitions
and theorems under their stated assumptions. It does not discharge the
\emph{Deployment obligations} or the \emph{External observation
assumption}, which a deployment must establish separately. The formal
development is publicly archived as
the Zenodo deposit \cite{andreakis2026dblog_virtual_cuts}
(\url{https://doi.org/10.5281/zenodo.21732790}). The archived
development declares no axioms. Every witness and counterexample
fixture is a constructed instance. The only type not introduced as a
datatype or record is a conservative typedef for source coordinates.

\smallskip
\noindent\textbf{AI assistance.} During the preparation of this work the author
used the generative AI assistants ChatGPT 5.5 (OpenAI)
and Claude Opus 4.7 (Anthropic) to support drafting and
revising the exposition and developing the Isabelle/HOL
formalization. The author directed the research, reviewed
and edited all resulting text, and checked every definition,
theorem statement, and proof. The author is the sole author
of this paper and takes full responsibility for its content. The
AI assistants are tools, not authors.

\bibliography{refs}

\end{document}

%% file: figures/fig01_theorem_ladder/fig01.tex

\centering
\begin{tikzpicture}[setA]

\tikzset{
  ladderChip/.style={layerChip, minimum width=11mm, minimum height=11mm},
  ladderSlab/.style={block, text width=82mm, minimum height=11mm,
                     align=left, inner sep=4pt, font=\sffamily\footnotesize},
  ladderNote/.style={font=\sffamily\scriptsize\itshape, text=gMid,
                     text width=44mm, align=left, anchor=west,
                     inner sep=0pt},
  panelHd/.style={font=\sffamily\bfseries\footnotesize, text=gDark,
                  anchor=west},
  cardTtl/.style={font=\sffamily\bfseries\footnotesize, text=gDark,
                  anchor=north west, text width=62mm, align=left,
                  inner sep=0pt},
  cardAtt/.style={font=\sffamily\scriptsize\itshape, text=gMid,
                  anchor=north west, text width=62mm, align=left,
                  inner sep=0pt},
  cardTxt/.style={font=\sffamily\scriptsize, text=gDark,
                  anchor=north west, text width=62mm, align=left,
                  inner sep=0pt},
  cardTxtG/.style={font=\sffamily\scriptsize, text=gMid,
                   anchor=north west, text width=62mm, align=left,
                   inner sep=0pt},
  chipNow/.style={draw=gDark, line width=0.5pt, rounded corners=2pt,
                  fill=gPale, font=\sffamily\bfseries\scriptsize,
                  text=gDark, inner xsep=4pt, inner ysep=1.6pt,
                  anchor=north west},
  chipFut/.style={draw=gLight, line width=0.5pt, rounded corners=2pt,
                  fill=white, font=\sffamily\bfseries\scriptsize,
                  text=gMid, inner xsep=4pt, inner ysep=1.6pt,
                  anchor=north west},
  cardBox/.style={draw=gMid, line width=0.6pt, rounded corners=2.5pt,
                  fill=gBg},
}

\def\rowGap{2.5mm}

\node[panelHd] (hdA) at (-5.5mm,10mm)
  {The theorem ladder: source-side core (Layers 0--4)};

\node[ladderChip] (c01) at (0,0) {0\,/\,1};
\node[ladderSlab,  right=1.5mm of c01, anchor=west] (s01) {%
  \textbf{Source histories, replay events, abstract DBLog runs.}
  Defines $\WellformedRun$ (source-log retention, frontier discipline,
  watermark consistency, chunk-read fidelity).};
\node[ladderNote,  right=2mm   of s01]              (n01)                      {%
  The wellformed-run model.\\
  WF carries the source-side observation\\
  obligations. The certificate's external-\\
  observation premise enters at Layer 3.};

\node[ladderChip,  below=\rowGap of c01.south, anchor=north] (c2)              {2};
\node[ladderSlab,  right=1.5mm of c2,  anchor=west]          (s2)              {%
  \textbf{Source-side virtual-cut soundness.}
  $\WellformedRun \Rightarrow$ replay at frontier equals source on the
  run scope.};
\node[ladderNote,  right=2mm   of s2]                        (n2)              {%
  Same assumptions as L0/1. No \emph{new}\\
  assumptions (external observation\\
  is already inherited).};

\node[ladderChip,  below=\rowGap of c2.south, anchor=north]  (c3)              {3};
\node[ladderSlab,  right=1.5mm of c3, anchor=west]           (s3)              {%
  \textbf{Accepted-certificate soundness.}
  $\Verify(C{,}E)=\Accept \,\land\, \FaithfulSourceObservation \,\land\,$
  checker substrate $\Rightarrow \VirtualCut(C, H)$.};
\node[ladderNote,  right=2mm   of s3]                        (n3)              {%
  Layer 2 $+$\\
  checker-substrate obligations $+$\\
  External observation assumption.};

\node[ladderChip,  below=\rowGap of c3.south, anchor=north]  (c4)              {4};
\node[ladderSlab,  right=1.5mm of c4, anchor=west]           (s4)              {%
  \textbf{Whole-table specialization.}
  $\VirtualCut(C, H) \,\land\, \WholeTableClaimScope(C, H, b) \Rightarrow$
  unrestricted equality on the table at the frontier.};
\node[ladderNote,  right=2mm   of s4]                        (n4)              {%
  Layer 3 $+$\\
  $\WholeTableClaimScope(C, H, b)$.};

\foreach \a/\b in {c01/c2, c2/c3, c3/c4} {
  \draw[-{Latex[length=5pt, width=4pt]}, gDark, line width=0.9pt]
    ([yshift=-0.2pt]\a.south)
    -- ([yshift=0.2pt]\b.north);
}

\coordinate (divL) at ($(c4.south west) + (-6.5mm, -7mm)$);
\coordinate (divR) at ($(n4.east |- divL) + (0.5mm, 0)$);
\draw[gLight, line width=0.4pt] (divL) -- (divR);

\node[panelHd, anchor=north west] (hdB) at ($(divL) + (0,-4mm)$)
  {Post-core extension map: independent extensions, \emph{not} ladder rungs};

\node[cardTtl, anchor=north west] (a-ttl) at ($(hdB.south west) + (1.5mm,-3.5mm)$)
  {Source-side virtual-cut algebra};
\node[cardAtt, anchor=north west] (a-att) at ([yshift=-0.8mm]a-ttl.south west)
  {Extends the Layer~2 virtual-cut theorem.};
\node[chipNow, anchor=north west] (a-c1) at ([yshift=-1.5mm]a-att.south west)
  {Machine-checked};
\node[cardTxt, anchor=north west] (a-d1) at ([yshift=-0.8mm]a-c1.south west)
  {Continuation across frontiers and restriction to a sub-scope.};
\node[chipFut, anchor=north west] (a-c2) at ([yshift=-1.6mm]a-d1.south west)
  {Future work};
\node[cardTxtG, anchor=north west] (a-d2) at ([yshift=-0.8mm]a-c2.south west)
  {Disjoint composition, concrete extended-certificate verification,
   and raw overlap/retry normalization.};

\node[cardTtl, anchor=north west] (b-ttl) at ($(a-ttl.north west) + (72mm,0)$)
  {Destination-side sink extensions};
\node[cardAtt, anchor=north west] (b-att) at ([yshift=-0.8mm]b-ttl.south west)
  {Extend the core's certified clean prefix with a separate sink model.};
\node[chipFut, anchor=north west] (b-c1) at ([yshift=-1.5mm]b-att.south west)
  {Future work, Non-claim};
\node[cardTxtG, anchor=north west] (b-d1) at ([yshift=-0.8mm]b-c1.south west)
  {ACK-only receipt and stream-prefix correctness. Applied destination
   state is a Non-claim.};
\node[chipFut, anchor=north west] (b-c2) at ([yshift=-1.6mm]b-d1.south west)
  {Future work};
\node[cardTxtG, anchor=north west] (b-d2) at ([yshift=-0.8mm]b-c2.south west)
  {Strong sink-state barrier convergence needs a fenced, idempotent
   sink contract and visible apply-barrier evidence.};

\begin{scope}[on background layer]
  \node[cardBox, fit=(a-ttl)(a-att)(a-c1)(a-d1)(a-c2)(a-d2),
        inner sep=4pt] {};
  \node[cardBox, fit=(b-ttl)(b-att)(b-c1)(b-d1)(b-c2)(b-d2),
        inner sep=4pt] {};
\end{scope}

\end{tikzpicture}

\caption{Theorem ladder and extensions. Layers~0--4 form the
source-side core, with each layer using the premises of the layers above
it. Continuation and scope restriction extend the Layer~2 virtual-cut
result. Disjoint composition, extended-certificate verification, and
overlap/retry normalization remain Future work. Destination-state
results require a separate sink model. A cut obtained from an accepted
certificate retains the certificate assumptions and faithful-source
observation premise.}
\label{fig:theorem-ladder}

%% file: figures/fig02_running_example_timeline/fig02.tex

\centering
\renewcommand{\ULthickness}{0.6pt}
\begin{tikzpicture}[setA, x=1cm, y=1cm]

\def\xCa{1.50}
\def\xCb{4.00}
\def\xCc{6.50}
\def\xCd{9.50}
\def\xCe{15.05}                   
\def\axisY{0}
\def\writeY{0.85}
\def\brackY{-0.30}
\def\brackLabelY{-0.60}
\def\obsY{-1.55}

\def\eqY{3.55}                    
\def\eqX{\xCd}                    

\node[font=\sffamily\Large, anchor=center] (eq) at (\eqX, \eqY) {$=$};

\node[blockFill, anchor=east, text width=29mm, align=center,
      left=4pt of eq.west, font=\sffamily\footnotesize]
  (lbox) {%
  \textbf{$\Apply(\CleanPrefix) \restrict K$}\\[1pt]
  \begin{tabular}{@{\ }r@{\,:\,}l@{\ }}
    100 & $3000$  \\
    200 & $12000$ \\
    300 & $3500$  \\
  \end{tabular}};

\node[block, anchor=west, text width=29mm, align=center,
      right=4pt of eq.east, font=\sffamily\footnotesize]
  (rbox) {%
  \textbf{$\Src(b_0, H, f) \restrict K$}\\[1pt]
  \begin{tabular}{@{\ }r@{\,:\,}l@{\ }}
    100 & $3000$  \\
    200 & $12000$ \\
    300 & $3500$  \\
  \end{tabular}};

\node[sublabelDim, anchor=south] at ($(lbox.north -| eq) + (0,1.5pt)$)
  {at frontier $f = c_4$, on scope $K = \{100, 200, 300\}$};

\coordinate (eqSouth) at ($(eq |- lbox.south) + (0, -2pt)$);
\draw[guide, line width=0.5pt]
  (\xCd, \axisY) -- (eqSouth);

\draw[axisLine] (-0.10, \axisY) -- (15.65, \axisY);

\foreach \x/\lab in {\xCa/c_1, \xCb/c_2, \xCc/c_3} {
  \draw[axisTick] (\x, \axisY+0.06) -- (\x, \axisY-0.06);
  \node[anchor=north, font=\sffamily\scriptsize, inner sep=1pt]
    at (\x, \axisY-0.06) {$\lab$};
}
\draw[axisTick, line width=0.8pt] (\xCd, \axisY+0.10) -- (\xCd, \axisY-0.10);
\node[anchor=north, font=\sffamily\scriptsize\bfseries, inner sep=1pt]
  at (\xCd, \axisY-0.08) {$c_4 = f$};

\node[sublabelDim, anchor=west, inner sep=0pt]
  at (15.80, \axisY) {source-log};

\foreach \x/\op/\key/\val/\tag in {%
  \xCa/Upd/100/$3000$/w_1,
  \xCb/Ins/300/$2500$/w_2,
  \xCc/Upd/200/$12000$/w_3,
  \xCd/Upd/300/$3500$/w_4%
} {
  \node[pillFill, anchor=south] (w-\tag) at (\x, \writeY+0.10) {%
    \strut\textbf{\op}~\key\\\,$\to$~\val\,};
  \draw[guide, line width=0.35pt] (\x, \axisY+0.07) -- (\x, \writeY+0.10);
}

\draw[braceMirror] (\xCa-0.65, \brackY) -- (\xCa+0.65, \brackY);
\node[sublabel, anchor=north] (chAlab) at (\xCa, \brackLabelY)
  {chunk $A$ at $c_1$};

\node[pill, anchor=north, text width=24mm, align=left, inner xsep=4pt]
  (chAobs) at (\xCa, \obsY) {%
  \,$100 \to 3000$\\
  \,\sout{$200 \to 10000$}};

\draw[braceMirror] (\xCc-0.65, \brackY) -- (\xCc+0.65, \brackY);
\node[sublabel, anchor=north] (chBlab) at (\xCc, \brackLabelY)
  {chunk $B$ at $c_3$};

\node[pill, anchor=north, text width=24mm, align=left, inner xsep=4pt]
  (chBobs) at (\xCc, \obsY) {%
  \,\sout{$300 \to 2500$}};

\draw[weakFlow, rounded corners=2pt]
  ($(chAobs.east) + (0.05, -0.10)$)
  to[out=15, in=-100, looseness=0.7]
  ($(w-w_3.south) + (-0.05, 0.02)$);
\draw[weakFlow, rounded corners=2pt]
  ($(chBobs.east) + (0.05, 0)$)
  to[out=15, in=-100, looseness=0.7]
  ($(w-w_4.south) + (-0.05, 0.02)$);

\node[font=\sffamily\scriptsize\itshape, text=gMid, anchor=north,
      inner sep=1pt, fill=white]
  at ($(w-w_3.south) + (0, -0.06)$) {dominated by};
\node[font=\sffamily\scriptsize\itshape, text=gMid, anchor=north,
      inner sep=1pt, fill=white]
  at ($(w-w_4.south) + (0, -0.06)$) {dominated by};


\draw[gLight, line width=0.4pt, densely dashed]
  (\xCe, \axisY) -- (\xCe, \eqY-0.45);

\node[pill, draw=gLight, text=gMid, anchor=south] (w-w5)
  at (\xCe, \writeY+0.10) {\strut\textbf{Del}~200\\\,$\to$~absent\,};
\draw[guide, gLight, line width=0.35pt]
  (\xCe, \axisY+0.07) -- (\xCe, \writeY+0.10);

\draw[axisTick, gMid, line width=0.8pt]
  (\xCe, \axisY+0.10) -- (\xCe, \axisY-0.10);
\node[anchor=north, font=\sffamily\scriptsize\bfseries, text=gMid,
      inner sep=1pt] at (\xCe, \axisY-0.08) {$c_5 = f'$};

\draw[braceMirror, draw=gLight, dashed]
  (\xCd+0.70, \brackY) -- (\xCe-0.70, \brackY);
\node[sublabelDim, anchor=north]
  at ($(\xCd, \brackLabelY)!0.5!(\xCe, \brackLabelY)$)
  {continuation segment: one CDC event};

\node[blockGhost, anchor=center, text width=28mm, align=center,
      font=\sffamily\scriptsize] (c5card) at (\xCe, \eqY) {%
  \textbf{cut advances to $f' = c_5$}\\[1.5pt]
  no new chunk read};

\end{tikzpicture}

\caption{Two chunk reads at different source-log coordinates assemble
into a replay whose state at frontier $f = c_4$ equals the source on
scope $K = \{100, 200, 300\}$. Stale observations (strikethrough) are
dominated by later source-log writes. No single physical read gathers
every key in scope, so the cut is virtual rather than physical.
That is the bootstrap cut at $c_4$. The lighter elements to the right
show the continuation of
Section~\ref{sec:advancing-cut-running-example}. One further
source-log event ($c_5$, a delete of account 200) is appended as the
continuation segment for the half-open interval $(c_4, c_5]$. It begins
strictly after the frontier $c_4$ and requires no additional chunk read. This advances
the same cut to the later frontier $f' = c_5$.}
\label{fig:running-example-timeline}

%% file: figures/fig03_dblog_mechanism/fig03.tex

\centering
\begin{tikzpicture}[setA, x=1cm, y=1cm]

\def\xCa{2.20}
\def\xCb{5.00}
\def\xCc{7.50}
\def\xCd{10.40}
\def\xWAlo{1.30}
\def\xWAhi{3.10}
\def\xWBlo{6.80}
\def\xWBhi{8.20}

\def\yCDC{1.30}
\def\yAxis{0}
\def\yRef{-1.45}

\draw[axisLine] (-0.10, \yAxis) -- (11.80, \yAxis);
\foreach \x/\lab in {\xCa/c_1, \xCb/c_2, \xCc/c_3, \xCd/c_4} {
  \draw[axisTick] (\x, \yAxis+0.07) -- (\x, \yAxis-0.07);
  \node[font=\sffamily\scriptsize, fill=white, inner sep=1pt]
    at (\x, \yAxis) {$\lab$};
}
\node[sublabelDim, anchor=west, inner sep=0pt, align=left]
  at (11.95, \yAxis) {source-log};

\node[sublabel, anchor=east, font=\sffamily\scriptsize\itshape, text=gMid,
      align=right, inner sep=0pt]
  at (-0.10, \yCDC) {CDC events\\(source writes):};

\foreach \x/\op/\key/\val in {%
  \xCa/Upd/100/3000,
  \xCb/Ins/300/2500,
  \xCc/Upd/200/12000,
  \xCd/Upd/300/3500%
} {
  \node[pillFill, anchor=south, font=\sffamily\scriptsize,
        inner xsep=3pt, inner ysep=1.5pt] (cdc-\key) at (\x, \yCDC-0.10) {%
    \strut\textbf{\op}~\key$\to$\val};
  \draw[guide] (\x, \yAxis+0.10) -- (\x, \yCDC-0.10);
}

\node[sublabel, anchor=east, font=\sffamily\scriptsize\itshape, text=gMid,
      align=right, inner sep=0pt]
  at (-0.10, \yRef) {Refresh events\\(chunk rows):};

\draw[braceMirror] (\xWAlo, \yAxis-0.20) -- (\xWAhi, \yAxis-0.20);
\node[sublabel, anchor=north, inner sep=2pt, fill=white]
  at ($(\xWAlo,\yAxis-0.40)!0.5!(\xWAhi,\yAxis-0.40)$)
  {chunk $A$};

\node[pill, anchor=north, font=\sffamily\scriptsize,
      inner xsep=3pt, inner ysep=1.5pt]
  (rA1) at (\xCa, \yRef+0.40)
  {\strut $A$:\,$100 \to 3000$};
\node[pill, anchor=north, font=\sffamily\scriptsize,
      inner xsep=3pt, inner ysep=1.5pt]
  (rA2) at (\xCa, \yRef-0.10)
  {\strut $A$:\,$200 \to 10000$};
\draw[guide] (\xCa, \yAxis-0.20) -- (\xCa, \yRef+0.40);

\draw[braceMirror] (\xWBlo, \yAxis-0.20) -- (\xWBhi, \yAxis-0.20);
\node[sublabel, anchor=north, inner sep=2pt, fill=white]
  at ($(\xWBlo,\yAxis-0.40)!0.5!(\xWBhi,\yAxis-0.40)$)
  {chunk $B$};

\node[pill, anchor=north, font=\sffamily\scriptsize,
      inner xsep=3pt, inner ysep=1.5pt]
  (rB1) at (\xCc, \yRef+0.20)
  {\strut $B$:\,$300$$\to$$2500$};
\draw[guide] (\xCc, \yAxis-0.20) -- (\xCc, \yRef+0.20);

\draw[-{Latex[length=5pt, width=4pt]}, gDark, line width=0.7pt]
  (5.85, \yRef-0.65) -- (5.85, \yRef-1.75)
  node[midway, fill=white, inner sep=2pt,
       font=\sffamily\scriptsize\itshape, text=gMid]
       {merged in source-log order};

\matrix[
  matrix of nodes,
  column sep=10pt,
  row sep=3pt,
  nodes={font=\sffamily\scriptsize, anchor=west, inner xsep=2pt, inner ysep=1pt},
  anchor=north,
  ampersand replacement=\&,
] (mr) at (5.85, \yRef-2.10) {%
  |[font=\sffamily\bfseries\scriptsize]| \# \&
    |[font=\sffamily\bfseries\scriptsize]| Kind \&
    |[font=\sffamily\bfseries\scriptsize]| Coord \&
    |[font=\sffamily\bfseries\scriptsize]| Event \&
    |[font=\sffamily\bfseries\scriptsize]| In normalized view? \\
  1 \& CDC \& $c_1$ \& Update~$100 \to 3000$        \& \checkmark \\
  2 \& Ref \& $c_1$ \& chunk $A$: $100 \to 3000$    \& \checkmark \\
  3 \& Ref \& $c_1$ \& chunk $A$: $200 \to 10000$   \& |[text=gMid, font=\sffamily\scriptsize\itshape]| dominated by 5 \\
  4 \& CDC \& $c_2$ \& Insert~$300 \to 2500$        \& \checkmark \\
  5 \& CDC \& $c_3$ \& Update~$200 \to 12000$       \& \checkmark \\
  6 \& Ref \& $c_3$ \& chunk $B$: $300 \to 2500$    \& |[text=gMid, font=\sffamily\scriptsize\itshape]| dominated by 7 \\
  7 \& CDC \& $c_4$ \& Update~$300 \to 3500$        \& \checkmark \\
};

\begin{scope}[on background layer]
  \node[draw=gDark, line width=0.5pt, rounded corners=1pt,
        inner sep=4pt, fit=(mr)] {};
\end{scope}

\draw[gMid, line width=0.35pt]
  (mr.west |- mr-1-1.south) ++(0, -1.5pt)
  -- (mr.east |- mr-1-1.south) ++(0, -1.5pt);

\end{tikzpicture}

\caption{The DBLog mechanism in the running example. Source
writes become \emph{CDC events} above the axis. Chunk-read rows become
\emph{refresh events} (below the axis, grouped by chunk and bracketed
by source-log watermarks). The two streams merge in source-log order
into the \emph{raw clean prefix} below, which is the list used by the
Layer~2 theorem. A refresh row whose key is touched by a later source
write is marked \emph{dominated}. The raw clean prefix keeps it, and an
optional normalized view may drop it without changing the $\Apply$
result (\S\ref{raw-replay-and-normalized-clean-prefix}).}
\label{fig:dblog-mechanism}

%% file: figures/fig04_virtual_vs_physical_cut/fig04.tex

\centering
\begin{tikzpicture}[setA, x=1cm, y=1cm]

\begin{scope}[xshift=0cm]

\node[font=\sffamily\footnotesize\bfseries, anchor=south]
  at (3.25, 3.95) {Physical cut};
\node[font=\sffamily\scriptsize\itshape, text=gMid, anchor=south]
  at (3.25, 3.50) {what DBLog \emph{does not} produce};

\draw[axisLine] (0.10, 0) -- (6.40, 0);

\def\xF{3.25}

\node[blockFill, anchor=center, align=center,
      font=\sffamily\scriptsize, inner xsep=5pt, inner ysep=3pt]
  (lstate) at (\xF, 2.30) {%
  all keys read at $f$\\[1pt]
  \begin{tabular}{@{\ }r@{\,:\,}l@{\ }}
    100 & $3000$  \\
    200 & $12000$ \\
    300 & $3500$  \\
  \end{tabular}};

\draw[gDark, line width=0.9pt] (\xF, 0) -- (lstate.south);
\node[font=\sffamily\scriptsize\bfseries, anchor=north, inner sep=1pt]
  at (\xF, -0.05) {$f$};

\node[font=\sffamily\scriptsize\itshape, text=gDark, anchor=north,
      align=center, text width=58mm]
  at (\xF, -0.90)
  {a single source-log coordinate at which\\
   every key in scope is observed};

\node[font=\sffamily\scriptsize, anchor=west, text=gMid,
      align=left, inner sep=1pt]
  at (\xF+0.20, 0.55)
  {monolithic\\\strut single-line cut};

\end{scope}

\draw[gMid, line width=0.5pt] (7.30, -1.30) -- (7.30, 4.10);

\begin{scope}[xshift=7.95cm]

\node[font=\sffamily\footnotesize\bfseries, anchor=south]
  at (3.25, 3.95) {Virtual cut};
\node[font=\sffamily\scriptsize\itshape, text=gMid, anchor=south]
  at (3.25, 3.50) {what DBLog \emph{does} produce};

\draw[axisLine] (0.10, 0) -- (6.40, 0);
\def\xCa{0.95}
\def\xCb{2.45}
\def\xCc{3.95}
\def\xCd{5.50}
\foreach \x/\lab in {\xCa/c_1, \xCb/c_2, \xCc/c_3} {
  \draw[axisTick] (\x, 0.06) -- (\x, -0.06);
  \node[font=\sffamily\scriptsize, anchor=north, inner sep=1pt]
    at (\x, -0.06) {$\lab$};
}
\draw[axisTick, line width=0.8pt] (\xCd, 0.10) -- (\xCd, -0.10);
\node[font=\sffamily\scriptsize\bfseries, anchor=north, inner sep=1pt]
  at (\xCd, -0.08) {$c_4 = f$};

\draw[braceMirror] (\xCa-0.30, -0.35) -- (\xCa+0.30, -0.35);
\node[font=\sffamily\scriptsize, anchor=north, inner sep=1pt]
  at (\xCa, -0.45) {chunk $A$};
\draw[braceMirror] (\xCc-0.30, -0.35) -- (\xCc+0.30, -0.35);
\node[font=\sffamily\scriptsize, anchor=north, inner sep=1pt]
  at (\xCc, -0.45) {chunk $B$};

\draw[guide] (\xCd, 0) -- (\xCd, 1.95);

\def\eqY{2.30}
\node[font=\sffamily\Large, anchor=center] (eq) at (3.25, \eqY) {$=$};
\node[blockFill, anchor=east, align=center,
      font=\sffamily\scriptsize, left=3pt of eq.west,
      inner xsep=5pt, inner ysep=3pt]
  (lbox) {%
  \textbf{$\Apply(\CleanPrefix) \restrict K$}\\[1pt]
  \begin{tabular}{@{\ }r@{\,:\,}l@{\ }}
    100 & $3000$  \\
    200 & $12000$ \\
    300 & $3500$  \\
  \end{tabular}};
\node[block, anchor=west, align=center,
      font=\sffamily\scriptsize, right=3pt of eq.east,
      inner xsep=5pt, inner ysep=3pt]
  (rbox) {%
  \textbf{$\Src(b_0, H, f) \restrict K$}\\[1pt]
  \begin{tabular}{@{\ }r@{\,:\,}l@{\ }}
    100 & $3000$  \\
    200 & $12000$ \\
    300 & $3500$  \\
  \end{tabular}};

\node[font=\sffamily\scriptsize, anchor=center, text=gMid,
      align=center, inner sep=1pt]
  at (3.25, 0.55)
  {\emph{no single line}\\ outcomes match at $f$};

\node[font=\sffamily\scriptsize\itshape, text=gDark, anchor=north,
      align=center, text width=58mm]
  at (3.25, -0.90)
  {chunk reads at distinct coordinates\\
   replay-at-frontier $=$ source-at-frontier on scope $K$};

\end{scope}

\end{tikzpicture}

\caption{Physical cut vs.\ virtual cut. \emph{Left:} a physical cut
requires a single source-log coordinate at which every key in scope is
observed, shown as a single vertical line. DBLog does \emph{not} produce
one. The values shown are the counterfactual outcome on $K$. \emph{Right:}
a virtual cut requires replay at the frontier to equal the source at
the frontier on the chosen scope. The chunk reads (as in
Figure~\ref{fig:running-example-timeline}) sit at distinct earlier
coordinates, but the replay reconstructs the source state at the frontier.}
\label{fig:virtual-vs-physical-cut}

%% file: figures/fig05_anchor_domain_table_scope/fig05.tex

\centering
\begin{tikzpicture}[setA, x=1cm, y=1cm,
    conjNum/.style={
      draw=gDark, line width=0.5pt, circle, fill=gPale,
      minimum size=4.5mm, inner sep=0pt, font=\sffamily\bfseries\scriptsize,
      text=gDark,
    },
    keyDot/.style={
      font=\sffamily\small\bfseries, text=gDark, inner sep=1pt,
    },
    ellOutline/.style={gDark, line width=0.5pt},
    ellName/.style={font=\sffamily\scriptsize\bfseries, text=gDark, inner sep=1pt},
  ]

\def\rectXmin{0}
\def\rectXmax{10.20}
\def\rectYmin{0}
\def\rectYmax{7.3}
\def\axisY{-1.30}

\draw[gDark, line width=0.6pt, rounded corners=3pt]
  (\rectXmin, \rectYmin) rectangle (\rectXmax, \rectYmax);

\node[font=\sffamily\bfseries\footnotesize, anchor=north west, inner sep=4pt,
      text=gDark]
  at (\rectXmin+0.12, \rectYmax-0.12) {$\Scope(C)$};

\def\anchorCx{3.7}
\def\anchorCy{4.85}
\def\touchedCx{6.5}
\def\touchedCy{4.85}
\def\ellRx{1.95}
\def\ellRy{0.85}

\node[ellName, anchor=south]
  at (\anchorCx-0.65, \anchorCy+\ellRy+0.10) {anchor-domain};
\node[ellName, anchor=south]
  at (\touchedCx+0.65, \touchedCy+\ellRy+0.10) {touched-between};

\begin{scope}
  \clip (\anchorCx, \anchorCy) ellipse [x radius=\ellRx, y radius=\ellRy];
  \fill[gPale] (\touchedCx, \touchedCy) ellipse [x radius=\ellRx, y radius=\ellRy];
\end{scope}

\draw[ellOutline]
  (\anchorCx, \anchorCy) ellipse [x radius=\ellRx, y radius=\ellRy];
\draw[ellOutline]
  (\touchedCx, \touchedCy) ellipse [x radius=\ellRx, y radius=\ellRy];

\node[keyDot] at (\anchorCx-1.15, \anchorCy)   {$100$};
\node[keyDot] at ($(\anchorCx, \anchorCy)!0.5!(\touchedCx, \touchedCy)$) {$200$};
\node[keyDot] at (\touchedCx+1.15, \touchedCy) {$300$};

\node[font=\sffamily\scriptsize\itshape, text=gMid, anchor=center,
      align=center, inner sep=2pt]
  at (5.10, 3.20)
  {union $=$ table-scope-between $= \{100, 200, 300\}$};

\def\replayCx{5.10}
\def\replayCy{1.55}
\def\replayRx{3.2}
\def\replayRy{0.70}

\node[ellName, anchor=south]
  at (\replayCx, \replayCy+\replayRy+0.10) {replay-keys};

\draw[ellOutline]
  (\replayCx, \replayCy) ellipse [x radius=\replayRx, y radius=\replayRy];

\node[keyDot] at (\replayCx-2.10, \replayCy) {$100$};
\node[keyDot] at (\replayCx,      \replayCy) {$200$};
\node[keyDot] at (\replayCx+2.10, \replayCy) {$300$};

\draw[axisLine] (0.10, \axisY) -- (10.10, \axisY);
\node[sublabelDim, anchor=west, inner sep=0pt]
  at (10.20, \axisY) {source-log};

\def\bX{2.60}
\def\fX{7.60}
\draw[axisTick] (\bX, \axisY+0.07) -- (\bX, \axisY-0.07);
\node[font=\sffamily\scriptsize, anchor=north, inner sep=1pt]
  at (\bX, \axisY-0.07) {$b$};
\draw[axisTick, line width=0.8pt] (\fX, \axisY+0.10) -- (\fX, \axisY-0.10);
\node[font=\sffamily\scriptsize\bfseries, anchor=north, inner sep=1pt]
  at (\fX, \axisY-0.07) {$f = \Frontier(C)$};

\node[font=\sffamily\scriptsize\itshape, text=gMid, anchor=south, inner sep=2pt]
  at ($(\bX, \axisY)!0.5!(\fX, \axisY) + (0, 0.30)$) {$b \le \Frontier(C)$};

\node[conjNum] (c1n) at (\bX-1.40, \axisY+0.45) {1};
\draw[depFlow, line width=0.4pt]
  (c1n.east) to[bend left=12]
  ($(\bX, \axisY)!0.5!(\fX, \axisY) + (0, 0.10)$);

\node[conjNum] (c2n) at (\rectXmin+0.65, \rectYmax-1.15) {2};
\draw[-{Latex[length=3pt, width=2.5pt]}, gMid, line width=0.4pt]
  (c2n.south east) to[bend right=10] ($(\anchorCx-\ellRx, \anchorCy+0.45)$);

\node[conjNum] (c3n) at (\rectXmax-0.65, \rectYmax-1.15) {3};
\draw[-{Latex[length=3pt, width=2.5pt]}, gMid, line width=0.4pt]
  (c3n.south west) to[bend left=10] ($(\touchedCx+\ellRx, \touchedCy+0.45)$);

\node[conjNum] (c4n) at (\rectXmax-0.65, \rectYmin+0.83) {4};
\draw[-{Latex[length=3pt, width=2.5pt]}, gMid, line width=0.4pt]
  (c4n.north west) to[bend right=10] ($(\replayCx+\replayRx, \replayCy-0.15)$);

\node[font=\sffamily\scriptsize, text=gDark, anchor=south east, align=right,
      inner sep=2pt]
  at (\rectXmax, \rectYmax+0.10) {%
  \textbf{$\WholeTableClaimScope(C, H, b) \;=\;
          (1) \land (2) \land (3) \land (4)$}};

\end{tikzpicture}

\caption{Anchor domain and whole-table claim scope. The four numbered
conjuncts of $\WholeTableClaimScope(C, H, b)$ are annotated on the
diagram. The base state $b_0$, fixed by context, is implicit in
$\WholeTableClaimScope$ and $\AnchorDomain$. \textbf{(1)} The anchor $b$ precedes the frontier on the
source-log axis: $b \le \Frontier(C)$.
\textbf{(2)} $\AnchorDomain(H, b) \subseteq \Scope(C)$.
\textbf{(3)} $\TouchedBetween(H, b, \Frontier(C)) \subseteq \Scope(C)$.
\textbf{(4)} $\ReplayKeys(\CleanPrefix(C)) \subseteq \Scope(C)$.
Concrete keys follow the worked example of Section~\ref{a-small-anchor-example}: $100$ sits in
anchor-domain only, $200$ in the overlap, $300$ in touched-between
only. Replay-keys are $\{100, 200, 300\}$.}
\label{fig:whole-table-claim-scope}